%% file: main.tex
\PassOptionsToPackage{table}{xcolor}
\documentclass[sigconf,nonacm]{acmart}

\usepackage{enumitem}
\usepackage{graphicx}
\usepackage{xcolor}
\usepackage{array}
\usepackage{bm}
\usepackage{multirow}
\usepackage{mdframed}
\usepackage[most]{tcolorbox}
\AtBeginDocument{%
  }

\begin{document}

\title{HiPaper: A Human-AI Collaborative System for Academic Literature Triage and Scholarly Sensemaking}

\author{Ye Jia}
\authornote{These authors contributed equally to this research.}
\email{ye-aimmeng.jia@connect.polyu.hk}
\affiliation{%
  \institution{The Hong Kong Polytechnic University}
  \city{Hong Kong SAR}
  \country{China}
}

\author{Pangjing Wu}
\authornotemark[1]
\email{pang-jing.wu@connect.polyu.hk}
\affiliation{%
  \institution{The Hong Kong Polytechnic University}
  \city{Hong Kong SAR}
  \country{China}
}

\author{Hayley Y.W. Gu}
\authornotemark[1]
\email{yanwugu@polyu.edu.hk}
\affiliation{%
  \institution{The Hong Kong Polytechnic University}
  \city{Hong Kong SAR}
  \country{China}
}

\author{Zhuohang Jiang}
\authornotemark[1]

\email{zhuohang.jiang@connect.polyu.hk}
\affiliation{%
  \institution{The Hong Kong Polytechnic University}
  \city{Hong Kong SAR}
  \country{China}
}

\author{Xu Yuan}
\authornotemark[1]
\email{xander.yuan@connect.polyu.hk}
\affiliation{%
  \institution{The Hong Kong Polytechnic University}
  \city{Hong Kong SAR}
  \country{China}
}

\author{Peter H.F. Ng}
\email{peter.nhf@polyu.edu.hk}
\affiliation{%
  \institution{The Hong Kong Polytechnic University}
  \city{Hong Kong SAR}
  \country{China}
}

\author{Zackary P. T. Sin}
\email{zackary-p-t.sin@polyu.edu.hk}
\affiliation{%
  \institution{The Hong Kong Polytechnic University}
  \city{Hong Kong SAR}
  \country{China}
}

\author{Qing Li}
\email{qing-prof.li@polyu.edu.hk}
\affiliation{%
  \institution{The Hong Kong Polytechnic University}
  \city{Hong Kong SAR}
  \country{China}
}

\author{Chen Li}
\authornote{Corresponding author.}
\email{richard-chen.li@polyu.edu.hk}
\affiliation{%
  \institution{The Hong Kong Polytechnic University}
  \city{Hong Kong SAR}
  \country{China}
}

\renewcommand{\shortauthors}{Jia et al.}

\begin{abstract}
Scholarly sensemaking requires researchers to iteratively retrieve, collect, and synthesize literature as their understanding evolves. However, existing scholarly tools provide only partial support for this process. Recent artificial intelligence (AI)-based systems generally rely on opaque submit-and-wait pipelines that it users’ ability to inspect, revise, and validate intermediate outputs. In this study, we present \textsc{HiPaper}, a human-AI collaborative literature review system informed by a formative study with academic experts. \textsc{HiPaper} organizes scholarly sensemaking as a three-stage interactive workflow spanning \textit{Retrieval}, \textit{Collection}, and \textit{Synthesis}, centralized on the human-in-the-loop paradigm that exposes intermediate AI-generated content for user inspection and intervention through a persistent workspace called \textit{Knowledge Pocket}. Through this workflow, users can iteratively refine search intent, curate and organize papers with AI-enriched metadata and clustering, and conduct evidence-traceable synthesis over selected literature. Our user study with 14 Ph.D. students showed that \textsc{HiPaper} efficiently supports scholarly sensemaking by enabling human-AI collaboration in both familiar and unfamiliar domains. Furthermore, the findings highlight how staged interactions with intermediate AI-generated content successfully scaffold user interactions throughout the sensemaking process.

\end{abstract}


\begin{teaserfigure}
  \centering
  \includegraphics[width=\textwidth]{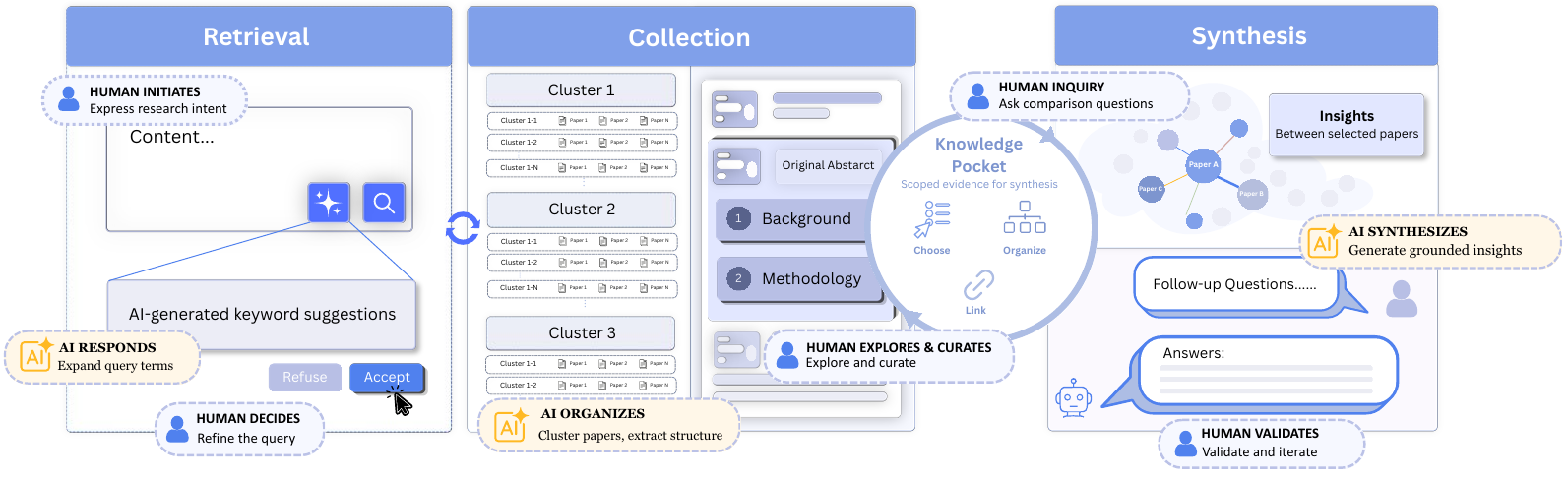}
  \caption{\textbf{\textsc{HiPaper}:} A system with three-stage human-AI collaborative workflow. \textit{Retrieval (Left):} User express research intent, and the AI expands query terms via keyword suggestions; users decide to accept or refuse. \textit{Collection (Middle):} AI organizes retrieved papers into clusters and extracts paper structures; users explore and curate evidence within a dedicated Knowledge Pocket workspace. \textit{Synthesis (Right):} Users ask comparison questions over curated papers; AI generates grounded insights; users validate and iterate through follow-up questions.}
  \Description{}
  \label{fig:HiPaper}
\end{teaserfigure}

\maketitle
\section{Introduction}
Scholarly sensemaking is a foundational research activity that helps scholars understand prior work, identify knowledge gaps, and develop new directions, primarily through literature review~\cite{2024Urquhart_Sense_making,uren2006sensemaking}. Researchers generally begin with vague ideas, refine their queries as they learn, compare findings across papers, and gradually construct an evolving understanding of a topic. Rather than consisting solely of paper retrieval, scholarly sensemaking unfolds as an iterative process of retrieval, collection, and synthesis~\cite{NIGHTINGALE2009381}, in which researchers refine queries, organize information, compare evidence, and build an evolving understanding of a topic.

Despite its importance, existing scholarly tools only partially support literature review as an iterative exploration process. Traditional scholarly retrieval systems, such as Google Scholar, primarily employ keyword-based paper retrieval and offer limited support for follow-up processes like comparing findings across papers. Recent generative artificial intelligence (GenAI)-based systems aim to automate the review process end-to-end via a ``submit-and-wait'' interaction model. Users specify an initial goal and passively wait while the system executes an opaque agentic pipeline, which can take up to 198 seconds to process one document~\cite{allabadi2025improving}. As a result, they create a fundamental mismatch with scholarly sensemaking, which develops through iterative refinement of search goals, evidence interpretation, and emerging understanding.

This mismatch is not merely a usability issue. When retrieval, collection, and synthesis are hidden inside an automated pipeline, users have limited opportunities to inspect intermediate outputs, revise search directions, or intervene before downstream synthesis is produced. Prior research suggests that such low-transparency interactions can frustrate users~\cite{gnewuch2022opposing} and may weaken their ability to critically assess evidence~\cite{bergstrom2022sense,bickmore2005establishing,Natarajan2025}. These limitations motivate a human-in-the-loop (HITL) approach to artificial intelligence (AI)-assisted scholarly sensemaking. Instead of hiding intermediate AI-generated content, such systems should expose it for user inspection and support actions such as accepting, rejecting, reorganizing, and building upon partial results across multiple stages of work. In this way, user steering can remain tightly coupled with AI assistance and better align with the iterative nature of scholarly sensemaking~\cite{jansen2025human}.

To ground this design space, we conducted a formative study with academic experts. Our findings revealed recurring concerns about the automated nature of current systems, including reduced control and the risk of hallucinated outputs. They also pointed to a three-stage review strategy: widespread literature retrieval, AI-assisted literature collection, and synthesizing evidence through focused cross-paper comparisons. These findings suggest that Human-AI collaboration for academic literature triage and sensemaking should be organized as a staged interaction workflow that supports sustained user control over intermediate stages.

Building upon these findings, we present \textsc{HiPaper}, a human-AI collaborative literature review system designed as a multi-stage HITL interactive workflow spanning \textit{Retrieval}, \textit{Collection}, and \textit{Synthesis}, as illustrated in Figure~\ref{fig:HiPaper}. \textsc{HiPaper} provides intermediate AI-generated content at each stage for user inspection and intervention. In \textit{Retrieval} stage, users draft search intent and refine it through AI-generated query suggestions. During \textit{Collection}, the system organizes retrieved papers into clusters and extracts paper structures, supporting user intervention through a dedicated structural workspace, termed ``Knowledge Pocket'', to store and organize selected papers. In \textit{Synthesis}, users ask paper-grounded questions over curated papers, and the system generates grounded insights that can be further probed through follow-up questions and validation.

\textsc{HiPaper} was evaluated through a user study with 14 Ph.D. students. Our findings show that \textsc{HiPaper} efficiently supports scholarly sensemaking by enabling human-AI collaboration across \textit{Retrieval}, \textit{Collection}, and \textit{Synthesis}. The study further illustrates how the system supports a staged human–AI collaborative workflow across both familiar and unfamiliar knowledge domains, enabling users to engage continuously with intermediate AI-generated content throughout the sensemaking process.

Our main contributions are summarized as follows:
\begin{enumerate}
\item We conducted a formative study to understand workflows in scholarly sensemaking, identifying a three-stage strategy and associated design needs for human-AI collaboration.
\item We present \textsc{HiPaper}, a human-AI collaborative system that combines a three-stage interaction workflow with \textit{Knowledge Pocket}, a persistent workspace for organizing intermediate outputs to support user inspection, curation, and validation throughout literature review.
\item Through a user study with 14 Ph.D. students, we showed that \textsc{HiPaper} efficiently supports scholarly sensemaking by enabling users to inspect, curate, and validate AI-generated intermediate outputs.
\end{enumerate}

\section{Related Work}

Scholarly sensemaking encompasses the iterative process of literature triage~\cite{van2024literature, farooq2017framework} and synthesis~\cite{knopf2006doing, nicholson2018introduction} across \textbf{Retrieval}, \textbf{Collection}, and \textbf{Synthesis}.

As a primary entry for literature retrieval, traditional academic search engines~\cite{kacperski2023examining,eger2025transforming}, such as Google Scholar~\cite{googlescholar}, Semantic Scholar~\cite{semanticscholar}, and DBLP~\cite{ley2002dblp}, still rely on keyword matching as a core mechanism, offering limited support for the complex cognitive activities involved in the literature review process~\cite{jordan2025ranking}. With the advancement of large language models (LLMs)~\cite{zhao2023survey,touvron2023llama,yang2025qwen3,jiang2025hibench}, recent research~\cite{yamada2025ai, he2025pasa} is exploring integrating agent-based capabilities into scientific discovery pipelines to transcend the limitations of traditional retrieval paradigms~\cite{ellis1998paradigms}. Although these approaches represent an advance over conventional search tools in terms of automation and reasoning, their underlying retrieval stage still largely relies on traditional search engines, leaving the fundamental limitations of keyword-based retrieval largely unaddressed.

The collection stage~\cite{mei2001use,wilcox2012research} refers to the phase in which researchers use reference management tools~\cite{zhang2012comparison} such as EndNote~\cite{endnote} and Zotero~\cite{zotero} to collect, filter, and organize the acquired literature after an initial retrieval, progressively building a more tractable and research-relevant corpus. Prior work~\cite{fahrudin2025improved,zhang2025paperbridge} has explored the development of intelligent systems to enhance both efficiency and automation in literature collection. Nevertheless, these approaches remain predominantly oriented toward collection and management efficiency, providing limited support for incremental sensemaking processes that emerge throughout the course of a literature review.

The synthesis stage~\cite{schick2016synthesis,suri2009advancements} refers to the process by which researchers integrate, interpret, and critically reflect on the collected literature, aiming to construct a coherent understanding of a given research domain. Related work~\cite{liu2022frontier,suh2023sensecape,shah2020prosumption} has explored methods such as visual organization to support researchers in progressively forming~\cite{ma2025garden} and evolving a holistic conceptual grasp of the field throughout the synthesis process. However, such approaches are highly dependent on individual researchers' expertise and manual intervention,  posing significant challenges in terms of both operational complexity~\cite{eimaraghy2004assessment} and time investment~\cite{rivera2007measuring}. Recently, GenAI tools~\cite{ren2025towards, reddy2025towards} such as Elicit~\cite{Whitfield03072023}, SciSpace~\cite{10.1007/978-3-031-56069-9_28}, and Research Rabbit~\cite{researchrabbit} have further expanded the modes of support available for literature synthesis. Nevertheless, these systems exhibit several critical shortcomings~\cite{bittle2025generative}, including limited process transparency, pronounced susceptibility to hallucination, and inadequate mechanisms for meaningful user intervention.

Relevance research~\cite {horvitz1999principles, amershi2019guidelines} indicates that more effective system design should reside in the collaborative zone between ``human-directed'' and ``fully automated'' approaches. Yet most existing tools remain confined to one of these two extremes, either manual retrieval~\cite{liu2022frontier,alfaruque2023integrating}  or AI-based generation~\cite{cai2025impact, 10.1145/3707640.3731913}, and consequently fail to support the full pipeline of the literature review process. There is a notable absence of systems that span the entire literature review workflow while substantially embodying the principles of HITL.

\section{Formative Study} \label{sec:formative}
We began by investigating how computer science researchers leverage AI-assisted tools for scholarly sensemaking, their typical workflows for this process, and the primary challenges they encounter. Consequently, these insights informed our design objectives.

\subsection{Methods}
\noindent\textbf{Prototype:} To effectively elicit insights from the experts, we developed an initial prototype to demonstrate the proposed interaction workflow. Drawing upon standard literature review practices and our own domain expertise, we implemented a paper retrieval interface, a dedicated view for displaying paper metadata, and an AI-powered conversational interface to support question-answering. This version of the system is shown in Appendix~\ref{prototype}.

\noindent\textbf{Participants:} Our formative study recruited five researchers at various career stages within the computer science domain. The cohort included two Ph.D. students (P1, P2), two postdoctoral fellows (P3, P4), and one Assistant Professor (P5). All participants possessed extensive research experience.

\noindent\textbf{Procedures:} Each participant was first asked to provide informed consent for the study. Subsequently, we conducted a semi-structured interview via online video conferencing software. Participants were asked a series of open-ended questions, such as: \textit{``How do you utilize AI-assisted tools when conducting literature reviews?''}, \textit{``Please describe your typical literature review workflow.''}, \textit{``What challenges do you face when using AI-assisted tools, and how do you resolve them?''}, \textit{``When writing a review, what is the primary difficulty you encounter in integrating scattered literature into a coherent narrative?''}, and \textit{``Overall, what would your ideal human-AI collaborative research assistant look like?''}

\noindent\textbf{Data Analysis:}
We used inductive thematic analysis to examine participants' practices, needs, and challenges. Two authors independently reviewed the interview transcripts and generated initial codes. They then met to discuss discrepancies, iteratively refine the codebook, and consolidate the codes into higher-level themes.

\subsection{Findings and Design Objectives}

Our findings are organized around three scholarly sensemaking stages, identified through thematic analysis. Based on the insights associated with each theme, we derive a corresponding design objective (DO) to inform the refinement of \textsc{HiPaper}.

\subsubsection{Retrieval: Intent Articulation, Search Quality, and Researcher Control}

Participants consistently raised concerns about the relevance and quality of AI-assisted search results. Current tools were seen as relying too heavily on keyword matching or superficial similarity, often producing mismatched results. As P5 noted, \textit{``search for literature based on keyword matching or keyword similarity... quality is uneven.''} P2 similarly described \textit{``It's not what I want... strange literature,''} while P1 observed that only \textit{``two or three of them will be what I'm really looking for.''} Participants also emphasized that retrieval should remain human-directed rather than fully automated. P4 stated that \textit{``humans [should] control which articles... are included in the analysis.''} P2 likewise argued that \textit{``the supervisor should dominate the research direction,''} and that researchers must \textit{``adjust and control the AI's direction in real time.''} P1 warned that otherwise, \textit{``the work that [the AI] delivers may not be really what you want... or even in the wrong direction.''}

Taken together, these findings suggest that researchers need retrieval support that better captures search intent, improves relevance, and preserves human control during iterative search.

\textbf{DO-R1:} The system should better capture users’ search intent and support iterative refinement of retrieval results for improved relevance and coverage.

\textbf{DO-R2:} The system should facilitate a HITL retrieval workflow by enabling researchers to inspect, review, and steer AI-assisted search outputs in real time.

\subsubsection{Collection: Efficient Triage, Metadata Support, and Lightweight Organization}

Participants consistently viewed literature triage as a human-directed process that AI should support rather than replace the human role in it. Researchers preferred screening papers manually using metadata such as titles, venues, keywords, and citations. As P4 noted, \textit{``I'll scan through all the titles... such as citations.''}. Participants also described lightweight evaluation and categorization practices during early triage. P3 explained this as \textit{``make a very preliminary classification... A, B, C, D [marks],''} and suggested that AI should assist rather than take over: \textit{``AI can provide a recommended score.''}

At the same time, participants highlighted the inefficiency of fragmented workflows and favored systems that preserve human agency while reducing manual overhead. P5 described \textit{``cross-platform writing manually... very difficult,''} and proposed \textit{``manual selection... categorizes, organizes, etc by AI.''} After viewing the prototype, participants further emphasized the value of high-signal metadata and concise summaries during triage. P3 preferred interfaces that \textit{``show the journal, its tier, and its citation count... the abstract is not the first requirement.''} P4 requested \textit{``one or two highly condensed sentences,''} and P1 noted that adding papers \textit{``one by one... [is] kind of annoying.''}

Taken together, these findings suggest the need for a unified collection space that supports human-directed paper triage through rich metadata, concise AI summaries, and lightweight organization.

\textbf{DO-C1:} The system should support efficient, human-directed triage by presenting rich metadata and concise AI summaries that help researchers rapidly assess paper relevance and quality.

\textbf{DO-C2:} The system should support lightweight organization of collected papers within a unified workspace, reducing cross-platform context switching while preserving user control over literature selection.

\subsubsection{Synthesis: Trust, Verification, Relationships, and Evidence-Grounded Sensemaking.}
A major concern raised by participants was the trustworthiness of AI-assisted literature review tools, especially during later-stage interpretation and synthesis. Participants emphasized that although AI can improve efficiency, its outputs must be carefully verified before they can be incorporated into real research workflows. For example, P1 noted, \textit{``I need to check both the author and the title... make sure they are correct.''} Participants also expressed concerns about the lack of transparency in how AI systems retrieve and prioritize information. As P3 explained, \textit{``it may have certain preferences and overlook some important information.''} These findings suggest that trust in AI-assisted literature review depends heavily on transparency and verifiability. Researchers need support for inspecting how results are generated and for validating AI outputs against source materials.

Participants raised strong concerns about the trustworthiness of AI-assisted literature review, especially during interpretation and synthesis. Although AI could improve efficiency, its outputs were seen as requiring careful verification. As P1 noted, \textit{``I need to check both the author and the title... make sure they are correct.''} Participants also worried about limited transparency in how AI retrieves and prioritizes information. As P3 explained, \textit{``it may have certain preferences and overlook some important information.''}
These findings suggest that synthesis support should be transparent, inspectable, and grounded in source materials. Participants also emphasized the need to understand both the thematic structure of a field and its development over time. For instance, participant P1 underscored the importance of \textit{``see[ing] the citation relationship between papers... connections between papers across different time periods,''} while P5 stressed the value of \textit{``creating a roadmap.''}

After viewing the prototype, participants further expressed a need for deeper synthesis support. P2 wanted to \textit{``understand the motivation behind researching these common problems.''} P5 asked to trace answers to \textit{``which paper and which specific section provides the evidence,''} and P4 emphasized \textit{``multiple classification systems... whether Paper A solves the limitations of Paper B.''}

In summary, participants need synthesis support to verify AI outputs, examine inter-paper relationships, and generate grounded interpretations with clear evidence traceability.

\textbf{DO-S1:} The system should support transparent, source-grounded validation of AI outputs, enabling researchers to inspect, verify, and trust the generated results during synthesis.

\textbf{DO-S2:} The system should provide multi-dimensional sensemaking support, allowing researchers to dynamically categorize literature, examine relationships between papers, and synthesize information with clear traceability to source materials.

\textbf{DO-S3:} The system should support focused cross-paper inquiry over user-selected papers, enabling researchers to iteratively ask questions, compare evidence, and refine emerging interpretations.

\begin{figure*} [h]
    \centering
    \includegraphics[width=0.85\linewidth]{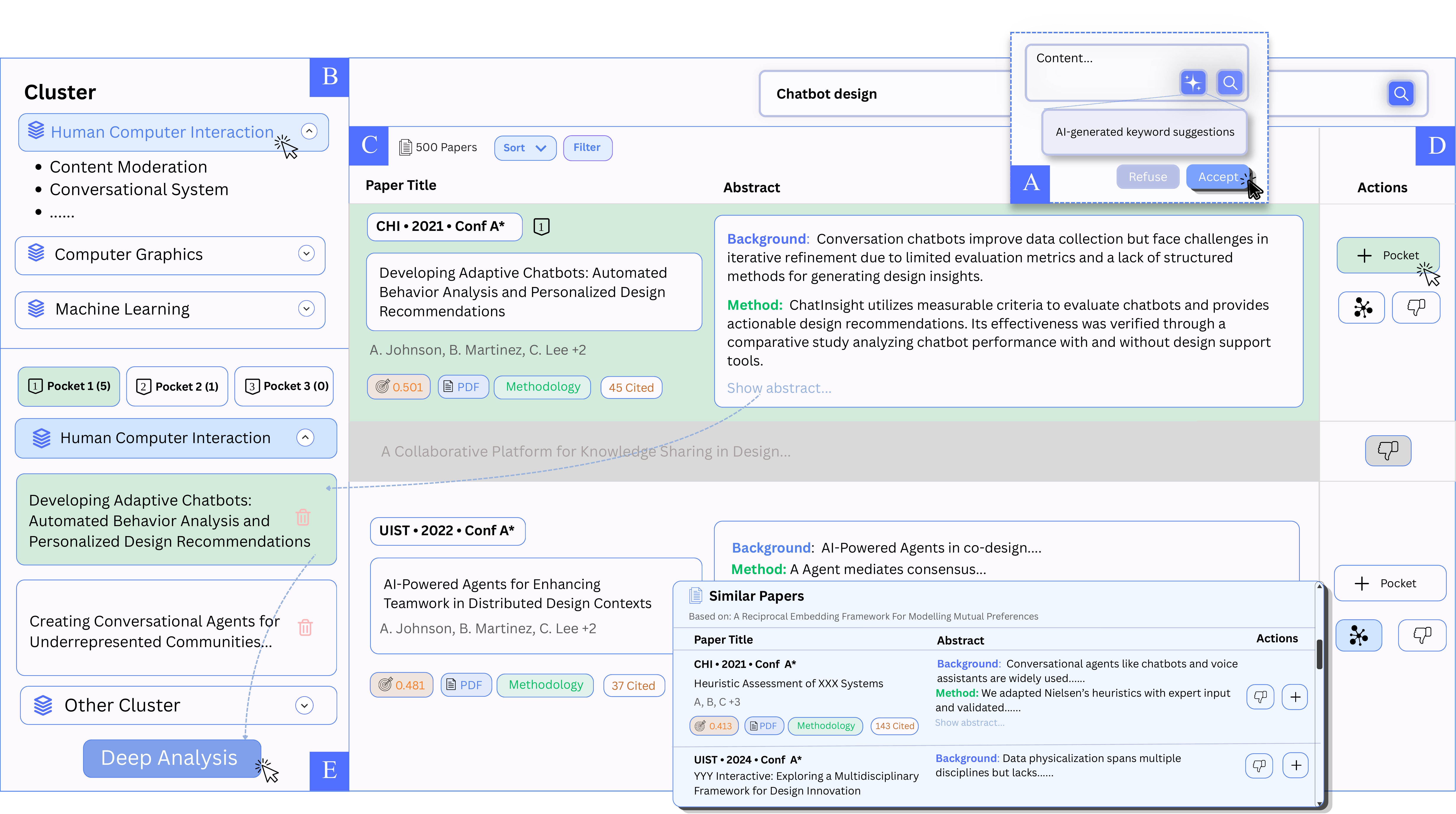}
    \caption{\textsc{HiPaper} retrieval and collection stages interface is organized into five panels. \textbf{(A) Search Panel:} Users enter a query; the AI expands the query with keyword suggestions, which users can Accept or Refuse via dedicated buttons; \textbf{(B)Cluster \& Pocket Panel: Retrieved papers are grouped into topic clusters (\emph{e.g.}, Human Computer Interaction, Machine Learning); users can organize curated papers into named Pockets (\emph{e.g.}, Pocket 1, Pocket 2) and trigger Deep Analysis on selected collections}; \textbf{(C) Paper List Panel:} each paper entry displays its title, authors, venue, year, conference tier, relevance score, and two AI-generated summaries — Background and Methodology — enabling rapid triage without opening the full paper; \textbf{(D) Action Panel:} Each paper row with three UI buttons including Add to Pocket, Similar paper, and Dislike. \textbf{(E) Knowledge Pocket Panel: Users continuously select papers from clusters and manage curated paper collections across multiple named Pockets.}}
    \label{fig:Searching_Collection}
\end{figure*}

\section{\textsc{HiPaper}}
Based on the identified DOs from the formative study, we present \textsc{HiPaper}, a human-AI collaborative system designed based on the HITL paradigm to support scholarly sensemaking, featuring three-stage LLM-powered functionalities for researchers. This section will demonstrate how the \textsc{HiPaper} assists researchers in improving literature triage and sensemaking through three iterative stages: \textit{Retrieval}, \textit{Collection}, and \textit{Synthesis}. The technical backbone supporting these functionalities is described in Appendix~\ref{sec:backend}.

\subsection{Retrieval}

To support users in expressing and refining their research intent (\textbf{DO-R1}) while preserving researcher control over AI-assisted retrieval (\textbf{DO-R2}), \textsc{HiPaper} provides an interactive retrieval workflow that combines AI assistance with user oversight.

\textsc{HiPaper} supports retrieval as an interactive process rather than a one-shot search action. Users can start with an initial topic or query and progressively refine their search direction as they receive results. This design addresses participants' concerns that existing retrieval tools often fail to adequately capture what they are actually looking for, especially when search intent is broad, evolving, or difficult to describe with only a few keywords.

To improve retrieval relevance and coverage, \textsc{HiPaper} incorporates AI support to interpret user intent beyond surface-level keyword matching. At the same time, the system does not fully automate retrieval decisions. Instead, users remain actively involved in reviewing returned papers, identifying mismatches, and steering the search toward more relevant directions. By allowing researchers to inspect and iteratively adjust the retrieval process, \textsc{HiPaper} supports a HITL workflow that better aligns retrieval outcomes with users' evolving goals.

This design addresses \textbf{DO-R1} by supporting iterative articulation and refinement of search intent, and \textbf{DO-R2} by enabling users to inspect, review, and steer AI-assisted retrieval outputs in real time.

\subsection{Collection}
The collection stage is designed to support efficient, human-directed triage of retrieved papers and lightweight organization of selected literature in a unified workspace, addressing \textbf{DO-C1} and \textbf{DO-C2}.

\textbf{AI Cluster (\textbf{DO-C2}).}
As illustrated in Panel B of Figure~\ref{fig:Searching_Collection}, the collection stage provides a cluster space that helps users organize retrieved papers into a structured topical landscape. Instead of relying solely on a flat ranked list, \textsc{HiPaper} groups papers into hierarchical clusters composed of root-level themes and more specific sub-clusters. This interactive design allows users to explore the literature at varying levels of granularity, inspect topical branches, and iteratively narrow their focus as their interests emerge. To support responsive interaction, cluster labels and related metadata are generated through an offline LLM-assisted preprocessing pipeline as described in section~\ref{sec:backend}.

\textbf{AI-Enriched Metadata (\textbf{DO-C1}).}
As shown in Panel C of Figure~\ref{fig:Searching_Collection}, \textsc{HiPaper} provides a set of enriched metadata cues to reduce users' effort in evaluating the quality and relevance of papers during collection. These cues include venue, publication year, relevance score, paper type, and citation count, which together support rapid triage and help users decide whether a paper should be collected for further inspection.

Under the \textit{Abstract} column in Figure~\ref{fig:Searching_Collection}, the system presents two AI-generated summaries by default: a concise description of the research background and a summary of the method. These summaries allow users to quickly assess a paper's focus and contribution before deciding whether to examine its full abstract.

\textbf{Paper Actions (\textbf{DO-C1}, \textbf{DO-C2}).}
\textsc{HiPaper} provides a dedicated quick-actions area for paper triage and curation (Panel D). After reviewing a paper's metadata, users can perform three primary actions. The first is \textit{Add to Pocket}, which is anticipated to be the most frequent action. Following Fitts' law~\cite{mackenzie1992extending}, this button has the largest visual area; clicking it gives immediate visual confirmation and adds the paper to the active Knowledge Pocket, as shown in Panel E of Figure~\ref{fig:Searching_Collection}. To minimize cognitive burden, Knowledge Pockets are purposely designed to be limited to three, aligning with research on working memory capacity~\cite{cowan2010magical}. Each pocket uses a hierarchical, cluster-based organization to clarify the structure of the collected papers.

The second action, \textit{Similar Paper}, opens a panel showing the top 10 related papers, allowing users to examine metadata and add papers to a pocket. The third action, \textit{Dislike}, collapses and grays out the paper entry, reducing visual salience to help focus on promising papers. After collecting a sufficient set, users click the \textit{Deep Analysis} button to proceed to the Synthesis Stage.

\subsection{Synthesis}
The synthesis stage aims to support transparent verification, multi-dimensional sensemaking, and focused cross-paper inquiry over user-selected literature, addressing \textbf{DO-S1}, \textbf{DO-S2}, and \textbf{DO-S3}.

\begin{figure*}[tbh]
    \centering
    \includegraphics[width=0.85\linewidth]{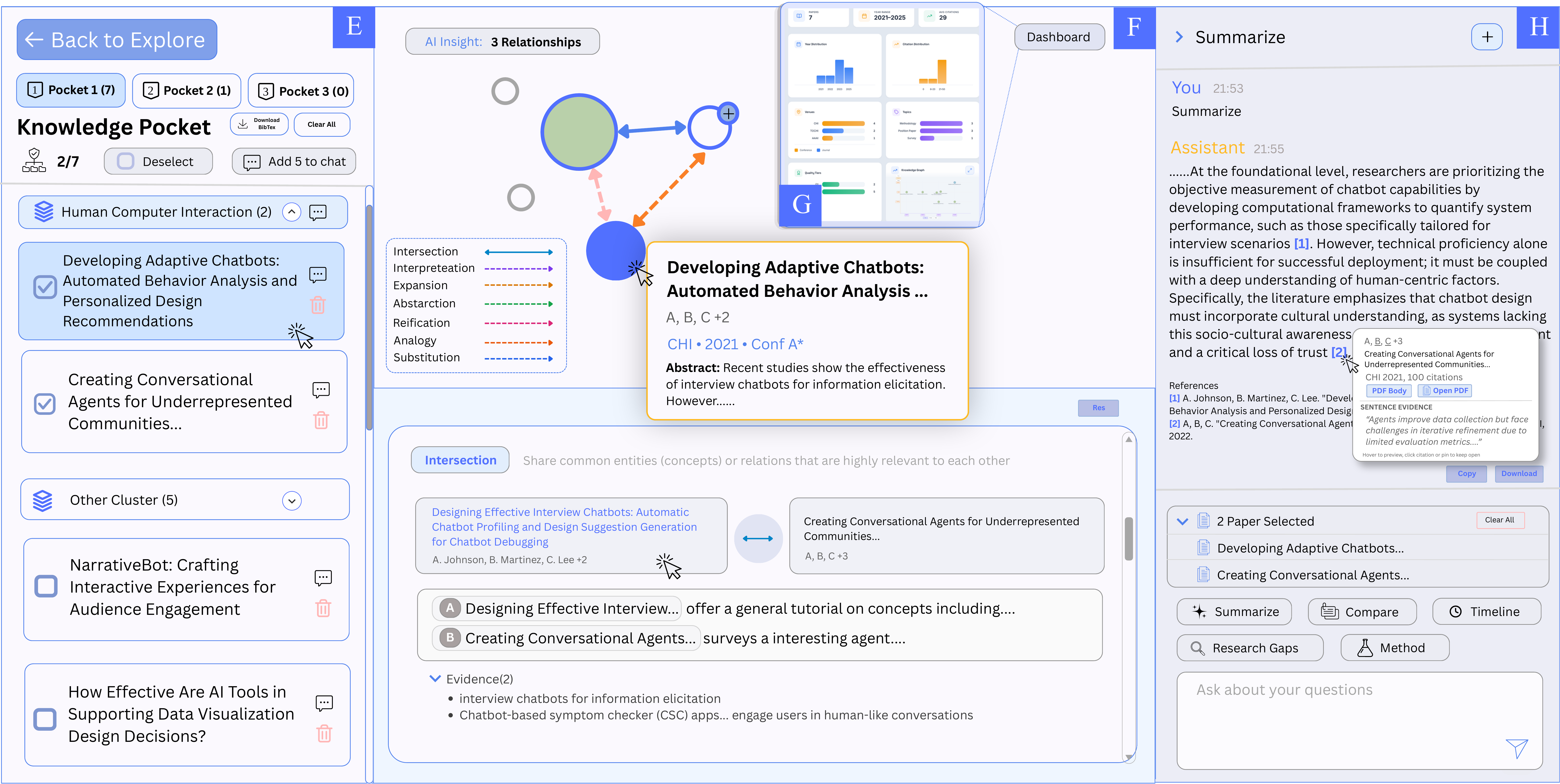}
    \caption{\textsc{HiPaper}'s synthesis interface, organized into four panels. (E) Knowledge Pocket Panel: Users manage curated paper collections across multiple named Pockets. Papers are organized by cluster, with checkboxes for selection, and options to Add to Chat; (F) AI Insight Dashboard: Displays an automatically generated visual summary of relationships (\emph{e.g.}, Different colored lines between nodes) among selected papers, including conceptual links such as Intersection, Interpretation, Expansion, Abstraction, Reification, and Analogy, helping users identify structural connections across literature; (G) Pocket Dashboard: analyzes papers in the Knowledge Pocket and presents a visual summary of structural connections among them. (H) Evidence-Traceable Chat Panel: Users trigger AI-generated synthesis actions, inlcuding \textit{Summarize}, \textit{Compare}, \textit{Timeline}, \textit{Research Gaps}, and \textit{Method} on selected papers. The AI produces grounded, citation-based responses with inline references (\emph{e.g.}, [1], [2]) traceable to specific papers. Users can further continue iterating via free-form questions in the input field.}
    \label{fig:UI_Function_Demostration}
\end{figure*}

\textbf{Knowledge Pocket as a Bridge (DO-S3):} As shown in Figure~\ref{fig:UI_Function_Demostration}, the \textit{Knowledge Pocket} serves as a bridge between Collection and Synthesis. Papers in Panel E curated during collection can be carried directly into the synthesis workspace and added to the chat panel (Panel H), enabling grounded analysis. By preserving user selections in a persistent workspace, the Knowledge Pocket supports a smooth transition from triage to deeper synthesis. A dashboard view of the knowledge pocket is provided in Panel G.

\textbf{AI Insight for Paper Relation Graph (DO-S1, DO-S2):} Also shown in Figure~\ref{fig:UI_Function_Demostration}, the synthesis stage provides an AI Insight Relation Graph to help users understand relationships among the literature they have collected (Panel F). Rather than requiring users to manually compare papers in sequence, \textsc{HiPaper} employed AI to get the potential inter-paper relationships based on the relationships among papers grounded on the prior research~\cite{doroudi2023related}, which contain relationships as \textit{Intersection}, \textit{Interpretation}, \textit{Expansion}, \textit{Abstraction}, \textit{Reification}, \textit{Analogy}, \textit{Substitution}, or \textit{No Meaningful Relationship}.

To support an interpretable and interactive synthesis process, the graph is generated progressively in a streaming manner, enabling users to observe how the relational structure emerges over time. When users click an edge, the system synchronizes the selected relationship with the relationship panel and presents an explanation of the inferred connection. Supporting evidence is also provided to help users validate the AI-generated interpretation. In addition, users can add papers represented in the graph directly to the chat panel for further discussion and analysis, creating a tighter loop between visual sensemaking and conversational synthesis.

\textbf{Evidence-Traceable AI Chat (DO-S3, DO-S1):}  To improve the robustness and trustworthiness of AI-assisted analysis, \textsc{HiPaper} incorporates a scoped-retrieval chat mechanism in the synthesis stage. Rather than generating responses over the entire corpus, the system constrains the AI's reasoning to papers that users explicitly add from the Knowledge Pocket and the paper relationship graph into the chat panel. This design helps keep the conversation grounded in user-selected literature and reduces the risk of irrelevant or unsupported responses. To further support verification, each AI response is accompanied by citations linked to the underlying paper content. When users hover over a citation marker, the system reveals the corresponding evidence, along with condensed paper metadata, in a floating window, enabling users to quickly trace and validate the source basis of the AI's answer.

\section{User Evaluation}
To evaluate \textsc{HiPaper}, we conducted a user study involving 14 Ph.D. students with backgrounds in computer science. Our primary research objectives were twofold: (1) to assess the overall usability of the system, and (2) to investigate how \textsc{HiPaper} supports researchers when navigating literature in both familiar and unfamiliar domains.

\textbf{Study Design.} Informed by the DOs derived from formative study, \textsc{HiPaper} is engineered to support researchers in exploring both familiar and unfamiliar research domains. To evaluate its efficacy, capability, and usability, we employed a within-subjects study design with counterbalanced task ordering. Participants were instructed to complete a rapid literature search task under two distinct conditions: one focusing on a topic within their existing area of expertise, and another exploring a research domain that was new to them.

\textbf{Participants.} This study recruited 14 participants from a single institution. All participants were doctoral students pursuing a Ph.D. degree and reported general research interests in computing-related fields. The participants comprised 10 men and 4 women, aged 25 to 30 years (\textit{M}~=~27.29, \textit{SD}~=~1.73). Because the study used a counterbalanced within-subjects design, participants were evenly assigned to one of two task orders, with 7 completing the unfamiliar-topic task first and 7 completing the familiar-topic task first. The study received ethical approval from the review board of [Institution Name] (reference: [IRB Number]).

\textbf{Experiment Task.} Because literature review inherently involves tracing idea origins, evaluating evidence, and identifying schools of thought~\cite{uren2006sensemaking}, we used it as the evaluation context for \textsc{HiPaper}. Based on our formative findings and prior work~\cite{pontis2016understanding}, we hypothesized that literature review behaviors would differ between familiar and unfamiliar topics.
To examine this, we designed a controlled task with an identical structure across conditions, varying only topic familiarity. In a pre-study questionnaire, participants rated candidate topics on a 5-point Likert scale. For each participant, a familiar topic (\(Topic_A\)) and an unfamiliar topic (\(Topic_B\)) were selected such that \(Score_{Topic_A} - Score_{Topic_B} \geq 3\). We also ensured that each topic returned more than 300 papers in direct search.

Specifically, the task is that participants were asked to identify three key themes in the assigned topic, select two to three representative papers for each theme with justifications, and briefly describe relationships among the themes within 25 minutes. To avoid forcing meaningless interaction, participants could stop once they felt they had completed the task. Notably, our goal was to analyze interaction patterns rather than review quality, which is difficult to assess objectively due to highly individualized review strategies. Finally, besides the time limit, we imposed no additional constraints on how participants used \textsc{HiPaper}.

\textbf{Procedure.} After registration, participants provided informed consent. \textit{1) Pre-questionnaire:} Participants reported demographics and topic expertise, which were used to assign tasks. \textit{2) Pre-study Interview:} Participants described their usual workflows and prior use of AI-based tools. \textit{3) Tutorial Session:} Participants watched a pre-recorded video introducing the main features of \textsc{HiPaper}. \textit{4) Task One:} To counterbalance order effects, half of the participants first completed a literature review task in a familiar domain and the other half in an unfamiliar domain. \textit{5) Post-questionnaire One:} Participants completed the NASA Task Load Index (NASA-TLX) and a customized User Experience Questionnaire (UEQ). \textit{6) Task Two:} Participants then completed a second task in the alternate domain. \textit{7) Post-questionnaire Two:} Participants completed the NASA-TLX, customized UEQ, and System Usability Scale (SUS). \textit{8) Post-study Interview:} Participants took part in a semi-structured interview about their experience using the system. \textit{9) End of Study:} Participants received a supermarket voucher worth approximately 13 USD as compensation for their time.

\textbf{Data Collection.}
We collected three types of data to evaluate \textsc{HiPaper}, including quantitative measures, qualitative feedback, and system interaction logs. Together, these data captured usability, perceived support for the literature review, and participants' actual behaviors during use.

\textit{Quantitative Data.}
We used standardized and customized instruments to assess participants' usability, workload, and user experience with the system. \textbf{General Usability} was measured using the SUS~\cite{brooke1996sus}. \textbf{Task Workload} was assessed using the NASA-TLX~\cite{hart2006nasa}. UEQ was measured with a customized multidimensional questionnaire on a 7-point Likert scale. The questionnaire covered four constructs, including \textit{Perceived Usefulness} (4 items, adapted from the Technology Acceptance Model~\cite{marangunic2015technology}), \textit{Search and Discovery Effectiveness} (3 items, based on Marchionini's Exploratory Search Theory~\cite{marchionini2006exploratory}), \textit{Organization and Sensemaking Support} (3 items, grounded in Sensemaking Theory~\cite{pirolli2005sensemaking,russell1993cost}), and \textit{Trust in AI Features} (7 items, developed from Human-AI Interaction guidelines~\cite{shneiderman2020human,amershi2019guidelines}).

\textit{Qualitative Data.}
We conducted semi-structured interviews before and after participants interacted with \textsc{HiPaper} to understand both their prior literature review practices and their experiences with the system. The pre-study interviews examined participants' existing workflows for literature review and scholarly sensemaking. The post-study interviews focused on their direct interactions with \textsc{HiPaper}, with attention to their perceptions, strategies, and overall evaluation of the platform.

\textit{User Behavior Logs.}
We also collected objective behavioral data through the built-in logging mechanism of \textsc{HiPaper}. The system recorded all user interactions, which were grouped into three stages. These included the \textit{Retrieval} stage, such as search button clicks and interactions with AI features in the search panel, the \textit{Collection} stage, such as interactions with Smart Clusters, paper management actions, and AI summarization features, and the \textit{Synthesis} stage, such as actions involving the AI insight panel, AI Chat, and the Knowledge Pocket.

\subsection{Behavior Log Process}

\textbf{Phase Progression Over Session Time.}
To characterize participants' progression through the sensemaking process, we mapped logged events to three high-level phases: \textit{Retrieval}, \textit{Collection}, and \textit{Synthesis}. For each session, time was normalized from \(0\%\) to \(100\%\) based on session start and end timestamps, allowing sessions of different durations to be compared on a common scale. We then estimated the proportion of sessions in each phase at each point of normalized session progress, which provides a dwell-weighted view of how activity phases evolve throughout a task session.

\textbf{Action-to-Action Transition Patterns.}
To capture participants' interaction strategies, we analyzed action-to-action transitions within each session. Since the number of logs across the two task conditions is uneven, we therefore use the possibility analysis rather than the raw event counts. We constructed transition matrices for familiar-topic and unfamiliar-topic sessions, converted them to conditional probabilities of the form \(P(j \mid i)\), and compared the two conditions using the difference \(P_{\mathrm{Fam}}(j \mid i) - P_{\mathrm{Unfam}}(j \mid i)\). This representation highlights which next-step actions were relatively more likely in one condition than the other, independent of overall differences in event volume.
Detailed processing steps, event mappings, and hyperparameters are provided in Appendix~\ref{appendix:log}.

\section{Results}
This section reports behavioral and quantitative results regarding the familiar and unfamiliar topic conditions.

\subsection{Behavioral Differences Between Familiar and Unfamiliar Topics}
\subsubsection{Phase Progression Over Session Time}
\begin{figure}
    \centering
    \includegraphics[width=\linewidth]{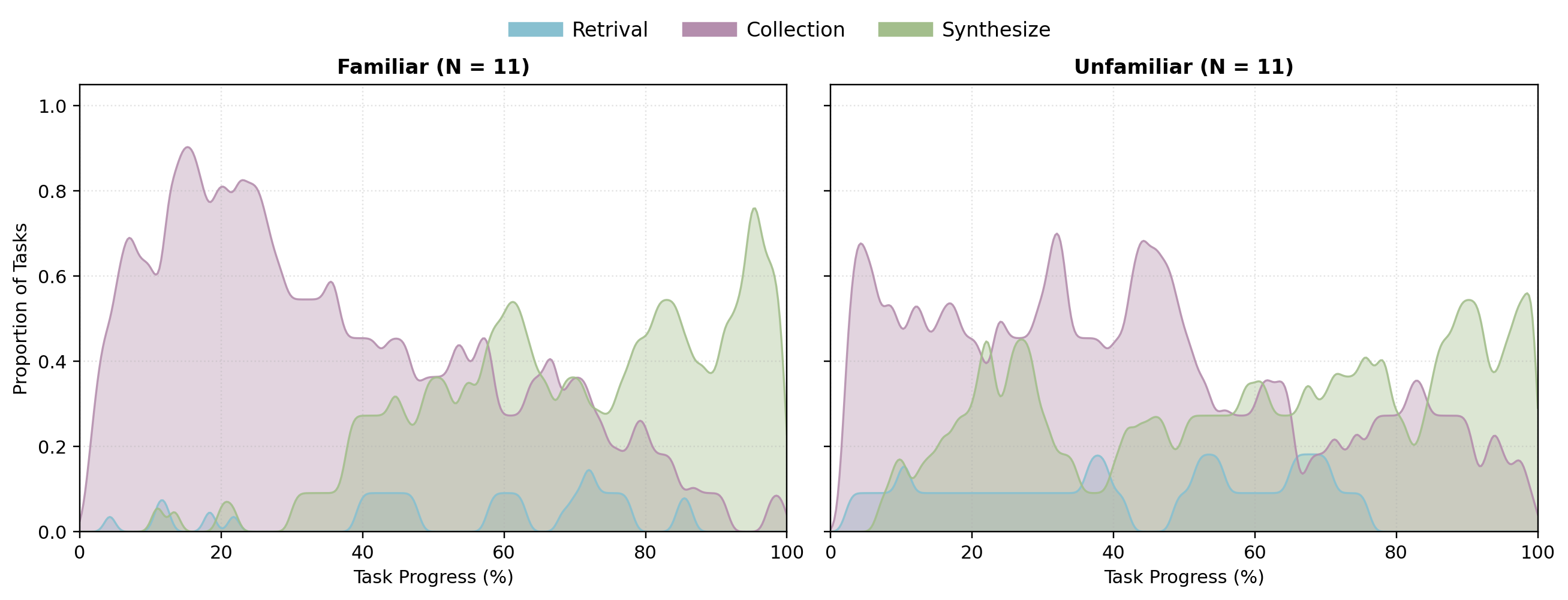}
    \caption{Phase Progression Over Task Time.}
    \label{fig:Phase_Progression}
\end{figure}

As illustrated in the Figure~\ref{fig:Phase_Progression}, the temporal distribution of the three states presents distinct patterns between the two types of tasks.
\textbf{Retrieval:} While accounting for a relatively small proportion overall, its distribution varied. In \textit{Familiar} tasks, retrieval was weak and scattered, appearing only occasionally in the middle and later stages. However, in \textit{Unfamiliar} tasks, retrieval was more intensive, forming distinct bands particularly during the 10\%-40\% and 50\%-70\% stages of the task progress. \textbf{Collection:} Participants under familiar topics demonstrated a typical front-focusing pattern. The collection held an absolute dominant position with a major peak, followed by a smooth, monotonic decrease until it almost completely disappeared at the end. In contrast, when participants were working on the unfamiliar topic, the Collection behavior persisted throughout the task. Although there were early peaks, this state maintained a high density and sharp fluctuations. \textbf{Synthesis:} For the participants under a familiar topic, Synthesis progressively increased starting from the middle stage and completely dominated the final stage of the task. However, when participants were under unfamiliar topics, Synthesis appeared fragmented. It emerged early in the task (at roughly 20\% progress) but fluctuated, resulting in a zigzag pattern that was highly entangled with the Collection curve throughout the middle and later stages.

\subsubsection{Action-to-Action Transition Patterns}
\begin{figure}[tp]
    \centering
    \includegraphics[width=\linewidth]{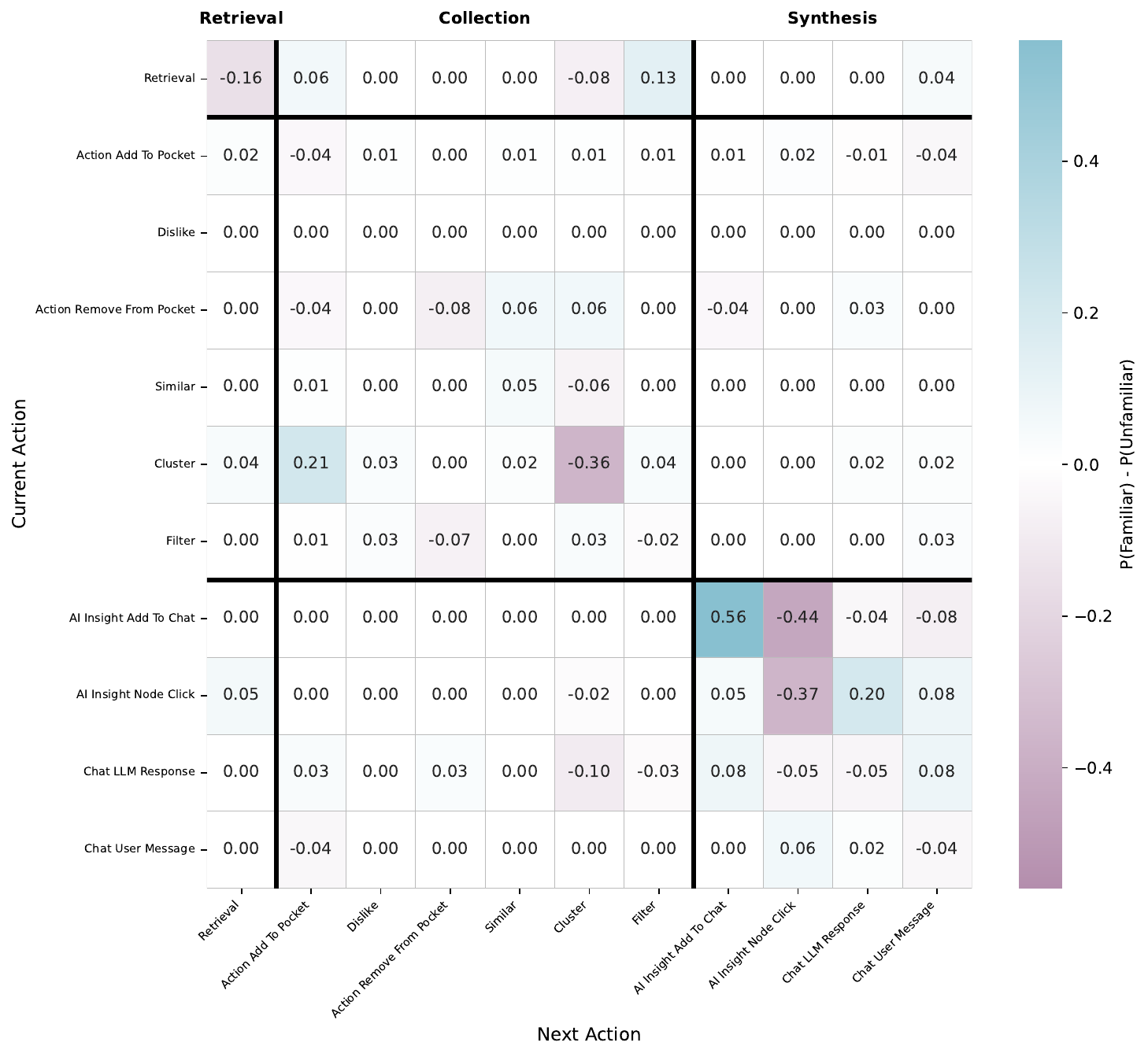}
    \caption{Action-to-Action Transition Heatmap.}
    \label{fig:a2a_heatmap}
\end{figure}

As shown in Figure~\ref{fig:a2a_heatmap}, Positive values indicate transitions that are more frequent in Familiar tasks, while negative values indicate transitions more frequent in Unfamiliar tasks.
\textbf{a) Retrieval:} Participants under Unfamiliar tasks display a higher frequency of consecutive Retrieval actions (-0.16) compared to Familiar tasks. \textbf{b) Collection:} Within the collection stage, Unfamiliar tasks are characterized by a higher frequency of repeated ``Cluster'' actions (-0.36). In contrast, Familiar tasks show a higher transition probability from ``Cluster'' to ``Action Add To Pocket'' (+0.21). \textbf{c) Synthesis:} The most pronounced differences occur within the synthesis stage. When participants are working on the Familiar, they exhibit a substantially higher probability of repeating the ``AI Insight Add To Chat'' action (+0.56). However, when they are under unfamiliar tasks,  a higher transition rate from ``AI Insight Add To Chat'' to ``AI Insight Node Click'' (-0.44), as well as repeated ``AI Insight Node Click'' actions (-0.37).

\subsection{Quantitative Results}
\noindent\textbf{Perceived Workload:}
A paired-samples t-test was conducted to evaluate the impact of task familiarity on users' perceived workload. The results indicated no statistically significant difference in the Overall NASA-TLX score between the Familiar $(M = 4.79, SD = 1.30)$ and Unfamiliar $(M = 5.19, SD = 1.45)$ conditions, $t(13) = 0.78, p = .452$. Similarly, no significant differences were found across the six individual subscales.

\noindent\textbf{User Experience:}
Paired-samples t-tests were conducted to compare user experience metrics between the familiar and unfamiliar task conditions. The analysis revealed a statistically significant difference in \textit{Search \& Discovery}; participants rated the system's support significantly higher in the unfamiliar condition ($M = 5.36$, $SD = 1.15$) compared to the familiar condition ($M = 4.83$, $SD = 0.98$), $t(13) = 2.35$, $p = .035$, Cohen's $d = 0.63$. No significant differences were observed for the remaining user experience dimensions. For \textit{Perceived Usefulness}, no significant difference was found (\(t(13) = 0.77\), \(p = .456\)) between the familiar (\(M = 5.29\), \(SD = 0.85\)) and unfamiliar (\(M = 5.45\), \(SD = 0.69\)) conditions. Similarly, \textit{Organization Support} showed no significant difference (\(t(13) = -1.21\), \(p = .247\)) between the familiar (\(M = 5.67\), \(SD = 0.78\)) and unfamiliar (\(M = 5.33\), \(SD = 1.00\)) conditions. For \textit{Trust in AI}, scores did not differ significantly (\(t(13) = 1.04\), \(p = .319\)) between the familiar (\(M = 5.04\), \(SD = 0.91\)) and unfamiliar (\(M = 5.24\), \(SD = 1.06\)) tasks. Finally, \textit{User Agency} also yielded no significant difference (\(t(13) = -1.24\), \(p = .239\)) when comparing the familiar (\(M = 5.57\), \(SD = 0.94\)) and unfamiliar (\(M = 5.21\), \(SD = 1.12\)) conditions. Additionally, the overall system usability, measured by the System Usability Scale (SUS) across all participants, achieved a mean score of 74.11 (\(SD = 11.66\)).

\noindent\textbf{Overall System Usability:}
Independent of the task conditions, the overall system usability was evaluated using SUS. Across all 14 participants, the system achieved a mean SUS score of 74.11 $(SD = 11.66$).

\section{Discussion}
Our findings indicate that AI-assisted scholarly sensemaking functions as an iterative, revisitable process rather than a rigid, linear progression from query to resolution. Both behavioral traces and interview data revealed that participants alternated continuously between the Retrieval, Collection, and Synthesis stages, particularly when navigating unfamiliar domains. Rather than indicating inefficiency, this non-linear workflow is a fundamental characteristic of AI-supported sensemaking; intermediate AI outputs frequently redefined participants' subsequent search strategies, relevance criteria, and overall task conceptualization.

To contextualize these quantitative and behavioral outcomes, we leverage participants' post-study interview feedback to elucidate how they utilized the system across both task conditions. Based on these insights, we derive four key implications for designing and evaluating interactive AI systems for scholarly sensemaking.

\subsection{Topic Familiarity Shaped Interaction Strategy}

The clearest pattern in our results is the contrast between familiar and unfamiliar tasks. In the phase progression analysis, familiar tasks showed a front-loaded Collection pattern followed by a relatively smooth transition into Synthesis. In contrast, unfamiliar tasks showed more sustained Retrieval activity and a more prolonged alternation between Collection and Synthesis. The customized UEQ results also suggested only \textit{Search \& Discovery} was rated significantly higher in the unfamiliar condition.

Participants' interviews help explain this difference. Several described unfamiliar-topic work as requiring a broad orientation before focused selection. For example, one participant explained that, in an unfamiliar domain, they first needed to understand ``what kinds of content exist at the level of broad modules'' before moving to more specific areas, whereas in a familiar domain they could \textit{``give more keywords from the beginning to find the subtopic [I] want (P4)''}. Another participant similarly noted that the system was \textit{``more helpful for the unfamiliar one, because it gives you an overview, while for familiar topics they instead wanted to ``go deeper into the papers (P3)''}. This interpretation is also echoed by participants who explicitly said that the system helped them \textit{``quickly understand a new field (P5)}'', but was less useful when they already had a strong mental model of the topic.

These findings suggest that the same interface mechanisms served different functions depending on prior knowledge. In familiar domains, participants used \textsc{HiPaper} more as a targeted triage and comparison tool. In unfamiliar domains, they relied on it more as an orientation scaffold. This complements prior critiques of one-size-fits-all literature tools by showing that the utility of retrieval, clustering, and synthesis aids is not fixed, but depends in part on the user's familiarity with the topic and resulting sensemaking needs~\cite{liu2022frontier,suh2023sensecape,shah2020prosumption}. For interactive AI systems, this suggests that there is no single best interaction mode for AI-assisted scholarly sensemaking. Systems may be more effective if they foreground different control surfaces depending on epistemic need, rather than offering uniform assistance across all review contexts.

\subsection{Literature Review with AI is Revisitable rather than Strictly Linear}

The findings of the user behavioral suggests that AI-supported scholarly sensemaking is iterative rather than one-pass. Especially for unfamiliar topics, participants moved repeatedly between retrieval, collection, and synthesis, with early synthesis often prompting further search and regrouping instead of ending the task. This resonates with classic accounts of sensemaking as iterative and evolving rather than sequential~\cite{2024Urquhart_Sense_making,uren2006sensemaking}.

Interviews reinforced this pattern. One participant described using the analysis view to first understand unfamiliar concepts and then returning to the retrieval stage: \textit{``when I don't know the concept ... I go to the analysis page ... use the AI to understand the definition and related concepts. Based on this understanding, I go back ... to find the paper'' (P7)}. Another participant described a similar progression from overview to targeted follow-up: they first used class-level summaries \textit{``to build a macro-level understanding, and then, when I wanted more detail about differences or relationships, I used AI chat to ask further questions (P4).''} These accounts align with the observed zigzag pattern in unfamiliar-topic sessions, where partial synthesis appeared to generate new insights rather than simply conclusions.

This suggests that the key design challenge is not only supporting each stage, but also enabling movement across stages without losing context. In \textsc{HiPaper}, artifacts such as clusters, selected papers, summaries, and pocketed papers persisted across phases and could be reused, helping participants build on partial understanding rather than restarting. This addresses a limitation in many existing systems, which tend to support retrieval, collection, or synthesis in relative isolation~\cite{zhang2012comparison,liu2022frontier,bittle2025generative}. We therefore argue that AI interfaces for scholarly sensemaking should support cross-phase revisitation rather than enforce a linear workflow.

\subsection{User-Curated Evidence Scopes Support Grounding and Control}

The transition heatmap further suggests that familiar tasks were more often moved from \textit{Cluster} to \textit{Add to Pocket}, whereas unfamiliar tasks were more likely to remain in repeated exploration. This move represents a concrete mechanism of interpretive control, situating the system within the collaborative zone between user direction and automation \cite{horvitz1999principles, amershi2019guidelines}. Furthermore, this behavior extends prior research on collection tools \cite{zhang2012comparison, fahrudin2025improved, zhang2025paperbridge} by demonstrating that user curation is not merely an organizational task but an epistemically vital grounding mechanism for AI-supported synthesis.

Interview data help explain why this happened. Participants frequently described the Knowledge Pocket as a way to turn candidate papers into a manageable evidence set. One participant said that using the three pockets to classify papers by different subtopics was \textit{``very useful to manage the papers and also make comparisons between these pockets (P1)'' }. Another participant explained \textit{``I used one pocket at a time for questions focused on a single topic, but added papers from multiple pockets when the question required comparison across topics (P12)''}. A third participant emphasized that paper selection itself should remain human-directed: \textit{``paper selection needs to be led by humans (P14)''}. These comments suggest that pocketing was not just a storage action. It marked a transition from open exploration toward bounded synthesis.

This is suggested in \textsc{HiPaper}, Knowledge Pockets were more than a bookmarking feature. Participants used them to assemble subsets of papers around a theme, compare papers within a set, and limit downstream analysis to papers they had explicitly selected. In this sense, pockets served as user-curated evidence scopes that shaped what the system could analyze at any given time. For design, this implies that systems should not only help users find papers, but also help them commit to a scoped subset that can be reused in downstream reasoning. In our data, this commitment appeared earlier and more directly in familiar-topic work, but the mechanism itself was valuable in both conditions.

\subsection{Intermediate Artifacts Function as Control Surfaces, not just Outputs}
Throughout the study, we observed participants treated AI-generated outputs not as static answers but as control surfaces to steer subsequent actions. This approach addresses critiques of GenAI transparency and hallucination \cite{bittle2025generative, ren2025towards, reddy2025towards} by facilitating meaningful user intervention. As P5 highlighted, manipulating these intermediate artifacts reduced manual labor while enhancing agency. By ensuring outputs remain inspectable and actionable, the system supports mixed-initiative interaction \cite{horvitz1999principles, amershi2019guidelines}, where AI complements scholarly judgment through human redirection and reinterpretation.

This instrumental approach clarifies why participants derived substantial value from the system, even while acknowledging the AI's inherent imperfections. Summaries, for instance, were predominantly deployed as triage instruments rather than substitutes for comprehensive reading. P4 utilized them to \textit{``make a preliminary judgment on whether a paper was actually worth a deeper dive,''} while P1 found that reviewing a \textit{summarized methodology was much easier than reading full abstracts''} for rapid literature filtering. Furthermore, when navigating unfamiliar domains, clustering provided a crucial high-level topography of the research landscape. Participant P4 noted that examining \textit{``broad categories first helped me quickly establish a macro framework before drilling down into specific sub-topics.''}

The relation graph likewise supported quick exploration of inter-paper connections. Participants used it after reviewing summaries or insights to ask follow-up questions and better understand relationships across papers. Participant P1 detailed: \textit{after reviewing initial summaries to find relevant topics, I would heavily rely on the chat AI to ask questions... to have a general understanding in this field.''} Participant P4 corroborated: \textit{I would simply ask the chat to expand on the AI's initial insights.''}

The key contribution of \textsc{HiPaper} is therefore not just integrating retrieval, clustering, and AI-chat, but enabling AI-generated and user-curated artifacts to persist across phases for continued reuse. More broadly, this suggests that interactive AI systems should treat outputs not only as answers, but as manipulable objects that support ongoing user control.

\subsection{Overall Usability was Promising, Long-Term Effects Remain Exploration}

The system's SUS score of 74.11 indicates reasonable usability, with participants like P4 praising the ``smooth and clear'' cohesive workflow. By consolidating search, management, and synthesis into a single human--AI pipeline, the system addresses the fragmentation typically observed in literature review processes \cite{zhang2012comparison, liu2022frontier, alfaruque2023integrating}. As P11 noted, this integration streamlines real-world workflows; however, our findings emphasize that such consolidation must be paired with the ability to inspect source documents and preserve scoped context to be fully effective.

At the same time, participants pointed out some disappointing design flaws. P5 expressed that \textit{``the most frustrating part was not being able to jump directly to the PDF''}. Others reported unclear graph references, such as \textit{```general labels in insight panel, like paper A and paper B'' } or \textit{``difficulty preserving the state of scoped context when reopening pockets''}. These frictions help explain why overall usability was positive but not yet excellent.

Additionally, because scholarly sensemaking is rarely a discrete, single-session effort, successful comprehension often requires the longitudinal tracking of a specific research topic. Consequently, the results of this study should be viewed as a preliminary exploration of the short-term effects of introducing a novel human-AI interaction paradigm for scholarly sensemaking.  This also leaves open an important gap relative to prior systems work, which has often evaluated literature tools through bounded tasks rather than sustained research practice~\cite{cai2025impact,10.1145/3707640.3731913}. The long-term impacts and sustained efficacy of this interaction paradigm remain a critical area for future longitudinal observation.

\section{Limitations}
There are several limitations to this study. The first limitation concerns the scope of the dataset; due to resource constraints, this study primarily collected literature published over the past five years, focusing on well-recognized venues within the HCI, Computer Graphics, and AI communities. Additionally, owing to copyright restrictions imposed by certain journals and conferences that lack Open Access policies, we restricted the literature data available in \textsc{HiPaper} to the abstract level. A second limitation is that both the formative and evaluation studies were conducted with participants possessing strong computer-related backgrounds. Consequently, the comprehensive evaluation of \textsc{HiPaper} among non-expert end-users remains an area for future exploration. The third limitation involves the loss of three sets of log data. Although we employed probability testing to mitigate this effect, we acknowledge that the missing data presents a limitation for analyzing user behavior logs. The last limitation lies in the study design. Because \textsc{HiPaper} adopts a HITL design, it is difficult to establish a fully fair comparison with either traditional scholarly retrieval tools, such as Google Scholar, or fully automated AI research tools, such as Deep Research. On the one hand, comparisons with conventional retrieval systems are complicated by the AI-assisted features embedded in \textsc{HiPaper}. On the other hand, comparisons with fully automated AI research tools are also imperfect, because the human involvement required by our system makes the final outcomes dependent on individual users. Moreover, the current study did not focus on assessing the quality of the final research outcomes themselves. We therefore acknowledge that direct comparative evaluation falls outside the scope of the present work and should be addressed in future long-term research.

\section{Conclusion}
In this study, we aimed to address the low transparency and fragmented workflow of AI-based scholarly sensemaking. Building on empirical findings from a formative study, we developed~\textsc{HiPaper}, a three-stage human-AI interaction workflow that transforms opaque LLM pipelines into manipulable interaction objects, exposing intermediate artifacts (\emph{e.g.}, paper selection, paper clusters, and grounded insights) for user inspection, curation, and validation across Retrieval, Collection, and Synthesis. Our user study with 14 Ph.D. students demonstrated that participants used the \textsc{HiPaper} system differently in familiar and unfamiliar domains. Overall, \textsc{HiPaper}'s three-stage AI-assisted workflow was perceived as an efficient and effective approach to scholarly sensemaking. Based on these findings, we identify several directions for future research: 1) Longitudinal studies are needed to examine real-world adoption over time. 2) Improve transparency, grounding, and trust in AI-generated synthesis. 3) Extending beyond computer science may help generalize \textsc{HiPaper}'s design principles to broader domains.

\bibliographystyle{ACM-Reference-Format}
\bibliography{bib}

\newpage

\clearpage
\appendix
\renewcommand{\thefigure}{\thesection.\arabic{figure}}
\setcounter{figure}{0}

\input{appendix/backend}
\input{appendix/prototype}
\input{appendix/data_process}
\input{appendix/dataset}
\input{appendix/prompt}

\end{document}

%% file: appendix/backend.tex
\section{Workflow of \textsc{HiPaper} Backbone} \label{sec:backend}

\begin{figure*}
    \centering
    \includegraphics[width=1\linewidth]{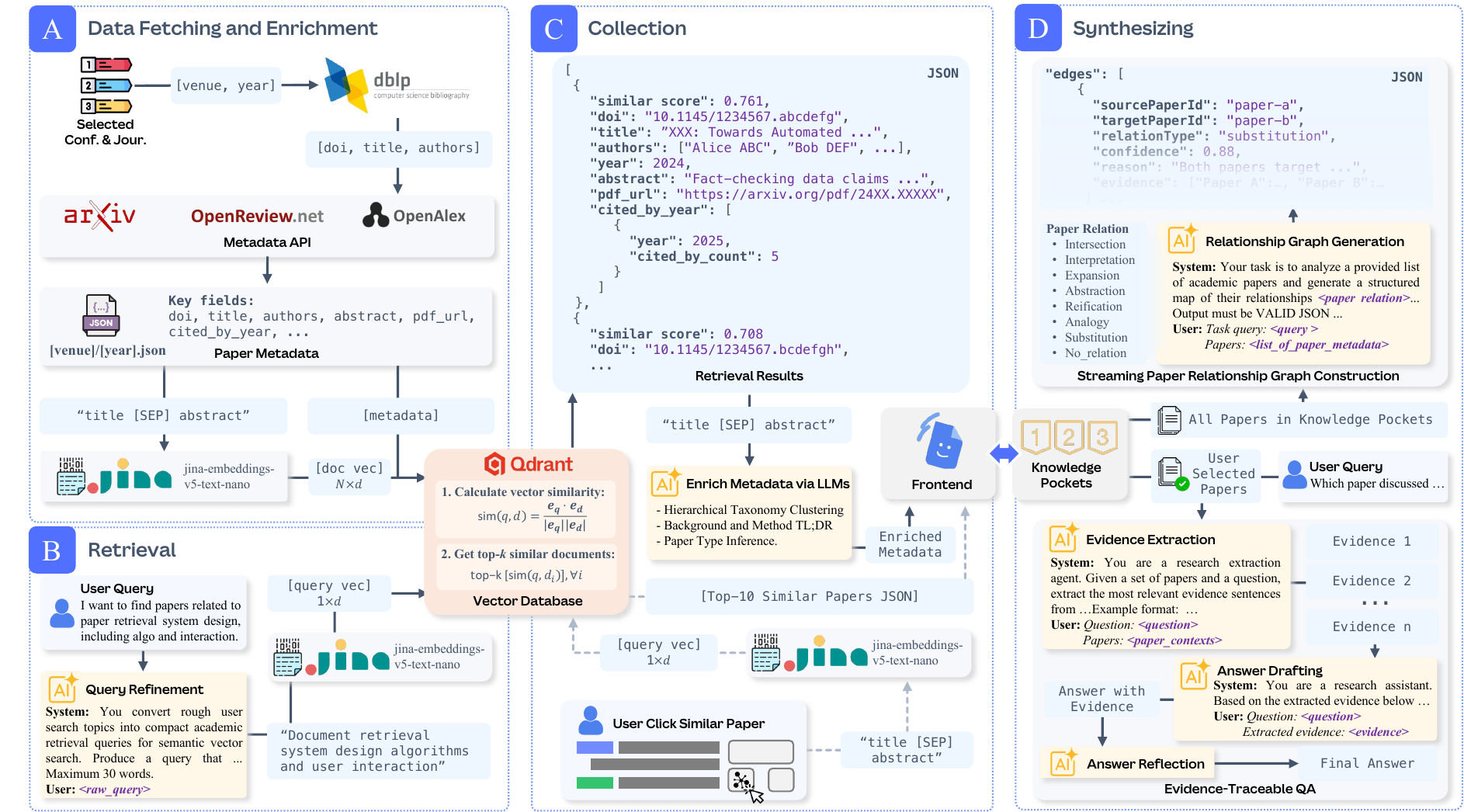}
    \caption{Backend workflow of \textsc{HiPaper}. The system moves from data fetching to retrieval, paper collection, and synthesis. (a) Papers from selected venues and years are fetched from metadata sources, normalized into a unified record format, and encoded from title and abstract for vector indexing. (b) A user’s rough query is rewritten into a retrieval-oriented academic query to retrieve semantically similar papers from the vector database. (c) Retrieved papers are collected as structured records and enriched with LLM-generated metadata. Users can also add similar papers through similarity-based browsing. (d) Selected papers are organized into \textit{Knowledge Pockets} for streaming paper relationship graph construction and evidence-traceable QA.}
    \label{fig:backend}
\end{figure*}

Figure~\ref{fig:backend} illustrates the overview of our backend system, revealing the implementation of each backend module that supports human-AI collaboration in academic literature triage and sensemaking.

\subsection{Data Fetch and Enrich}
We curate our corpus with 131,715 papers from 39 high-impact computer science conference and journal venues that primarily cover HCI, AI, Graphics, and Databases, with papers published from 2021 onward, as listed in Table~\ref{tab:venue-summary} (Appendix~\ref{app:datasets}). We first use the DBLP API~\cite{ley2002dblp} to \textbf{fetch} papers for each venue-year pair and extract core metadata to yield a reliable and de-duplicated paper set. We then query OpenAlex~\cite{priem2022openalex}, arXiv~\cite{arxiv}, OpenReview~\cite{openreview}, and paper landing pages to \textbf{enrich} complementary metadata such as abstracts, PDF links, and references with lightweight quality checks. All fields are normalized into a unified schema, as illustrated in Table~\ref{tab:data-schema} (Appendix~\ref{app:datasets}).

\subsection{Retrieval}
Traditional academic search often relies on keyword matching, which works when users know the exact terminology. However, it can miss semantically relevant papers phrased differently. To enhance retrieval performance, we adopt \textbf{Semantic Retrieval} to represent queries and papers in a shared embedding space, enabling semantic retrieval. Specifically, We encode each paper's title $t_d$ and abstract $a_d$ with a state-of-the-art text embedding model \texttt{jina-embeddings-v5-text-nano}~\cite{jina_embeddings_v5} as
$\mathbf{e}_d = \mathrm{Embed}(t_d, \texttt{[SEP]}, a_d)$.
It produces a 768-dimensional dense vector, which we store in \texttt{Qdrant}~\cite{qdrant} for efficient vector similarity search. At retrieval time, the user query is encoded with the same model to obtain $\mathbf{e}_q$, and papers are ranked by cosine similarity,
$\mathrm{sim}(q,d) = \frac{\mathbf{e}_q \cdot \mathbf{e}_d}{\|\mathbf{e}_q\|\|\mathbf{e}_d\|}$.
The system returns the top-$k$ papers, with default $k=500$, enabling retrieval of semantically similar work even without exact keyword overlap. Since users are often more familiar with keyword-style search, their initial queries are usually non-specific for semantic retrieval. We therefore add an LLM-based \textbf{Query Refinement} module that interprets user intent and rewrites the input into a more descriptive academic query. The prompt for query refinement is provided in Appendix~\ref{app:prompts}.

\subsection{Collection}
To support users to rapidly select preferred papers, our system enriches papers with a hierarchical research taxonomy, concise background and method summaries, and a paper type label. We propose a three-stage LLM pipeline with human verification to address the inadequacy of existing taxonomies for this setting, since curated systems (e.g., ACM CCS~\cite{acm_ccs}) evolve slowly and miss emerging terms, while automated taxonomies~\cite{priem2022openalex} yield uneven hierarchies, overlapping categories, and unstable boundaries. \textbf{i)} In the generation stage, the model assigns each paper a taxonomy from its title and abstract, including one research field and method topics. \textbf{ii)} In the reflection stage, the model revises duplicated concepts, hierarchy errors, and inconsistent granularity in its own outputs. We use \texttt{Qwen-3-4B-Instruct}~\cite{yang2025qwen3} for these two stages to support efficient large-scale processing. \textbf{iii)} In the combination stage, \texttt{GPT-5.4}~\cite{gpt5} merges candidate taxonomies, resolves inconsistent depths, and normalizes semantically similar categories. Human experts then review and revise the final hierarchy. In addition, We prompt LLMs to provide further information to help user capture the main idea of papers by briefly \textbf{summarizing papers' backgrounds and methods} and \textbf{inferring paper types}. Detailed prompts are provided in Appendix~\ref{app:prompts}.

\subsection{Synthesis}
After collecting the preferred papers into the knowledge pockets, our system conducts \textbf{Paper Relationship Generation} and provides \textbf{Evidence-Traceable QA} to assist users in the synthesis step of the literature review.

\paragraph{Streaming Paper Relationship Generation}
To help users capture relationships between papers, we implement a \textbf{Streaming Paper Relationship Generation} that incrementally extracts semantic edges from paper titles and abstracts. Given selected paper metadata, the system constructs a relation-extraction prompt based on a predefined relation schema~\cite{doroudi2023related}. The LLM then streams its output, while the system accumulates partial responses, parses complete relation edges as they become available, and renders them immediately. The front end updates the graph state online, progressively adding newly generated nodes and links. The prompt is provided in Appendix~\ref{app:prompts}.

\paragraph{Evidence-Traceable QA}
Accurate interpretation of scientific literature requires precise and verifiable statements. To reduce the risk of LLM hallucination, we ground responses in evidence extracted from the selected paper abstracts and add a final verification step. The evidence-traceable QA first identifies the source statements that are most relevant to the user's question via \textbf{Evidence Extraction}. These statements are then used as the only support for \textbf{Answer Drafting}, and citation markers link each claim to its source passage. A final \textbf{Answer Reflection} prompt checks whether each claim is grounded in the cited evidence and removes or revises unsupported content. This design improves provenance and reduces unsupported claims. The prompt is provided in Appendix~\ref{app:prompts}.

%% file: appendix/prototype.tex
\section{Prototype}
\label{prototype}

\setcounter{figure}{0}

\begin{figure*}[t]
    \centering
    \includegraphics[width=0.8\textwidth]{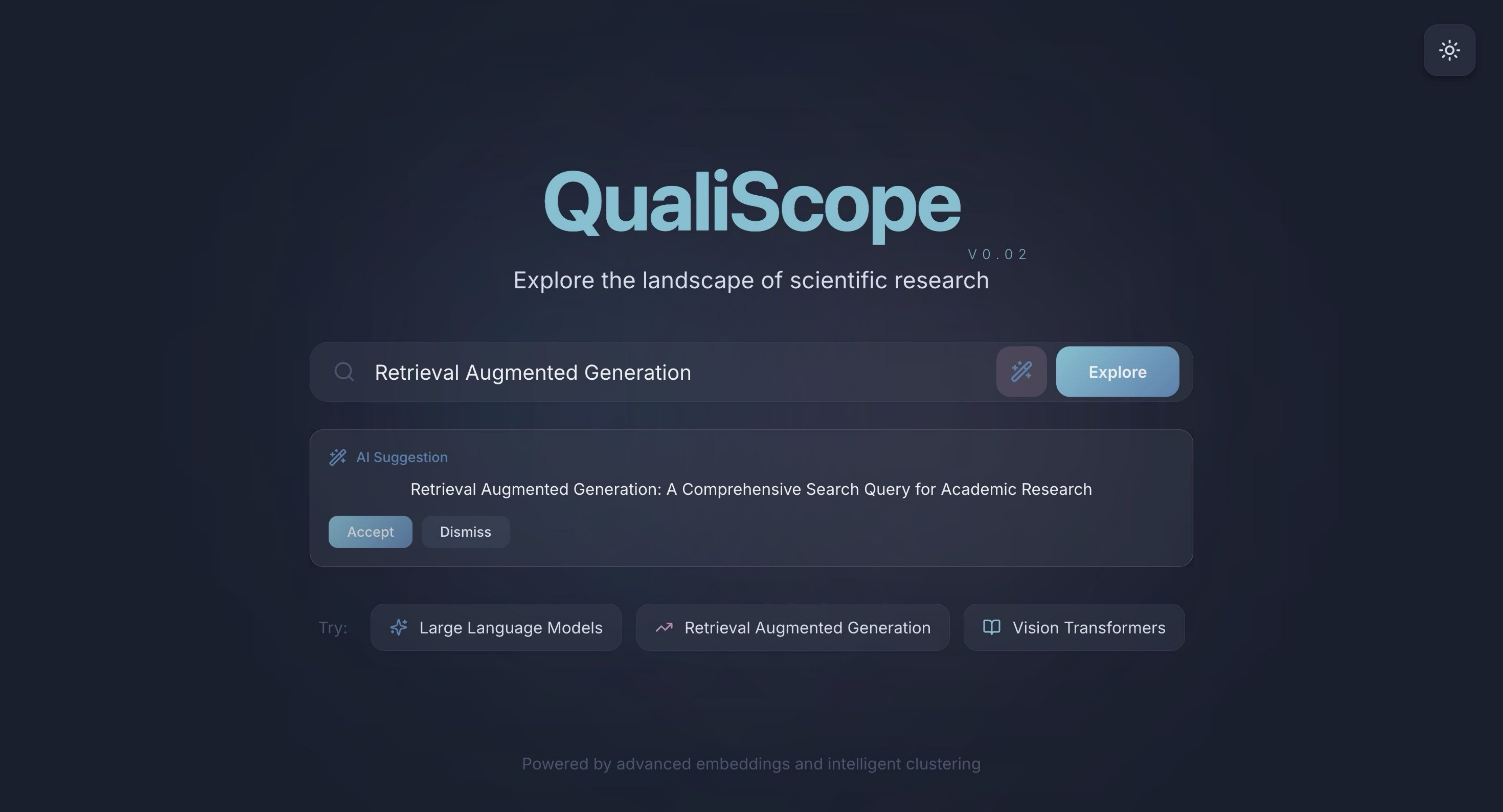}
    \caption{Prototype Retrieval Workspace.}
    \label{fig:prototype1}
\end{figure*}

\begin{figure*}[t]
    \centering
    \includegraphics[width=0.8\textwidth]{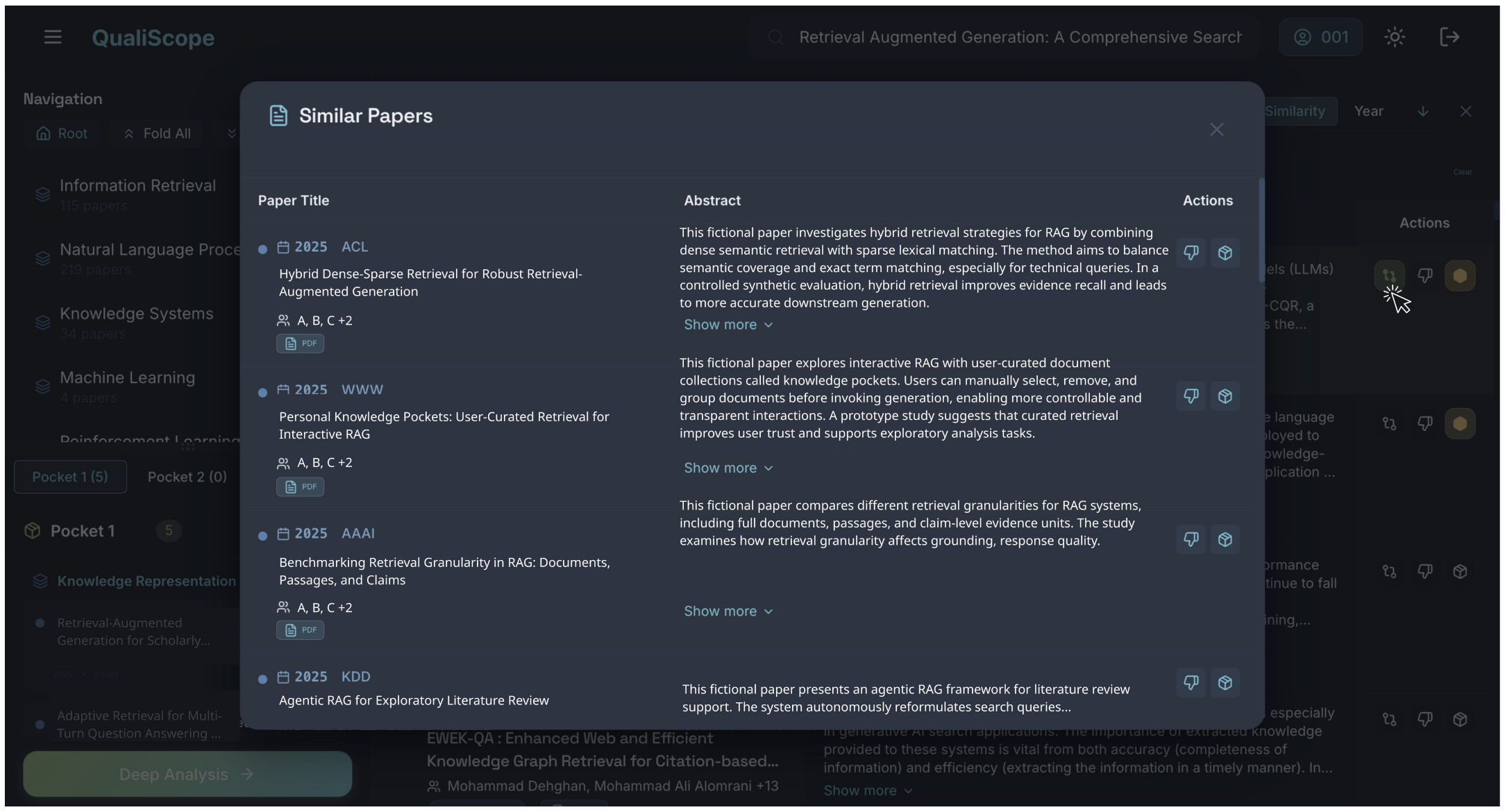}
    \caption{Prototype Collection Workspace.}
    \label{fig:prototype2}
\end{figure*}

\begin{figure*}[t]
    \centering
    \includegraphics[width=0.8\textwidth]{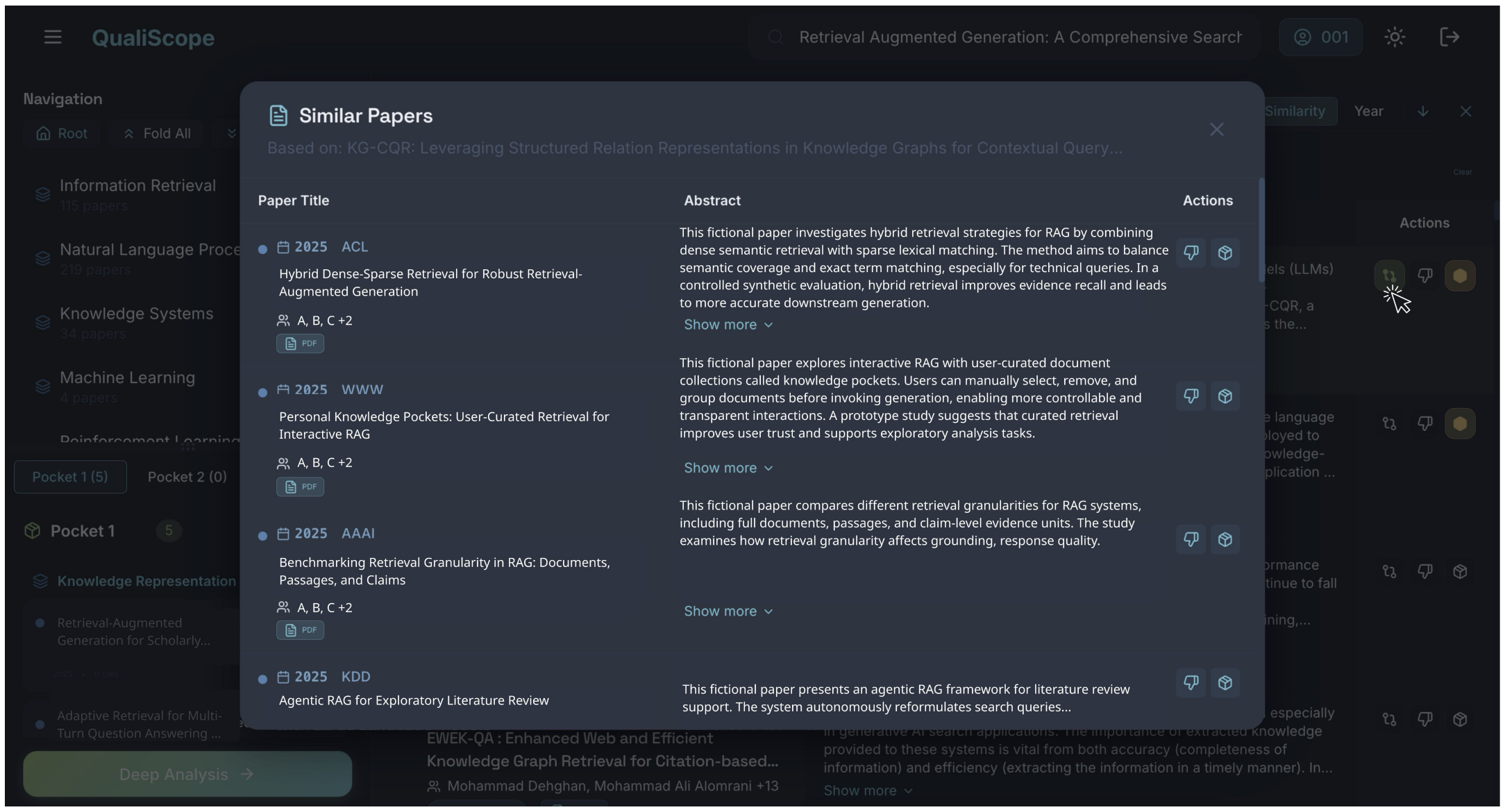}
    \caption{Prototype Similar-Paper Recommendation in the Collection Workspace.}
    \label{fig:prototype3}
\end{figure*}

\begin{figure*}[t]
    \centering
    \includegraphics[width=0.8\textwidth]{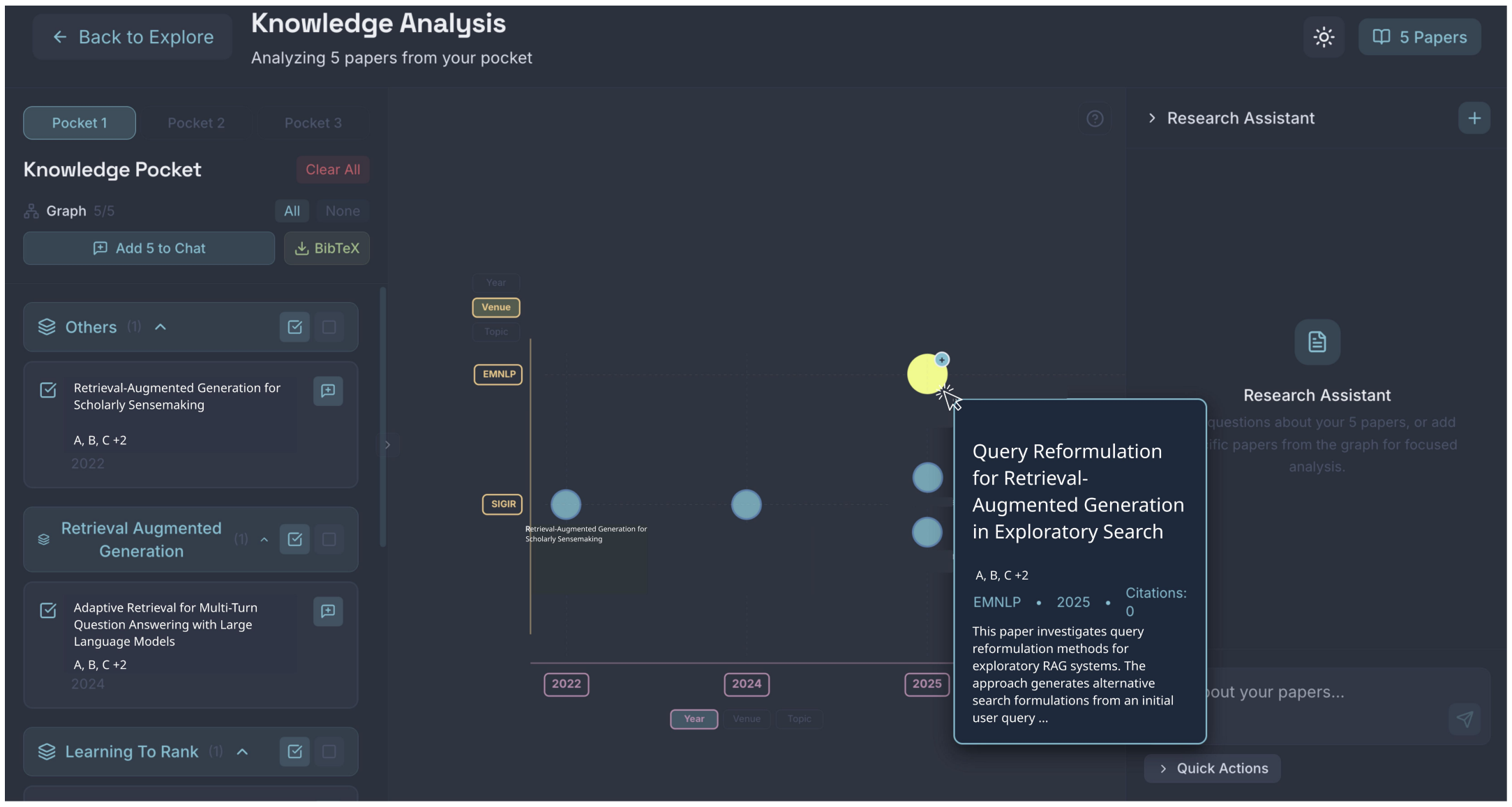}
    \caption{Prototype Synthesis Workspace for Paper Analysis.}
    \label{fig:prototype4}
\end{figure*}

\begin{figure*}[t]
    \centering
    \includegraphics[width=0.83\textwidth]{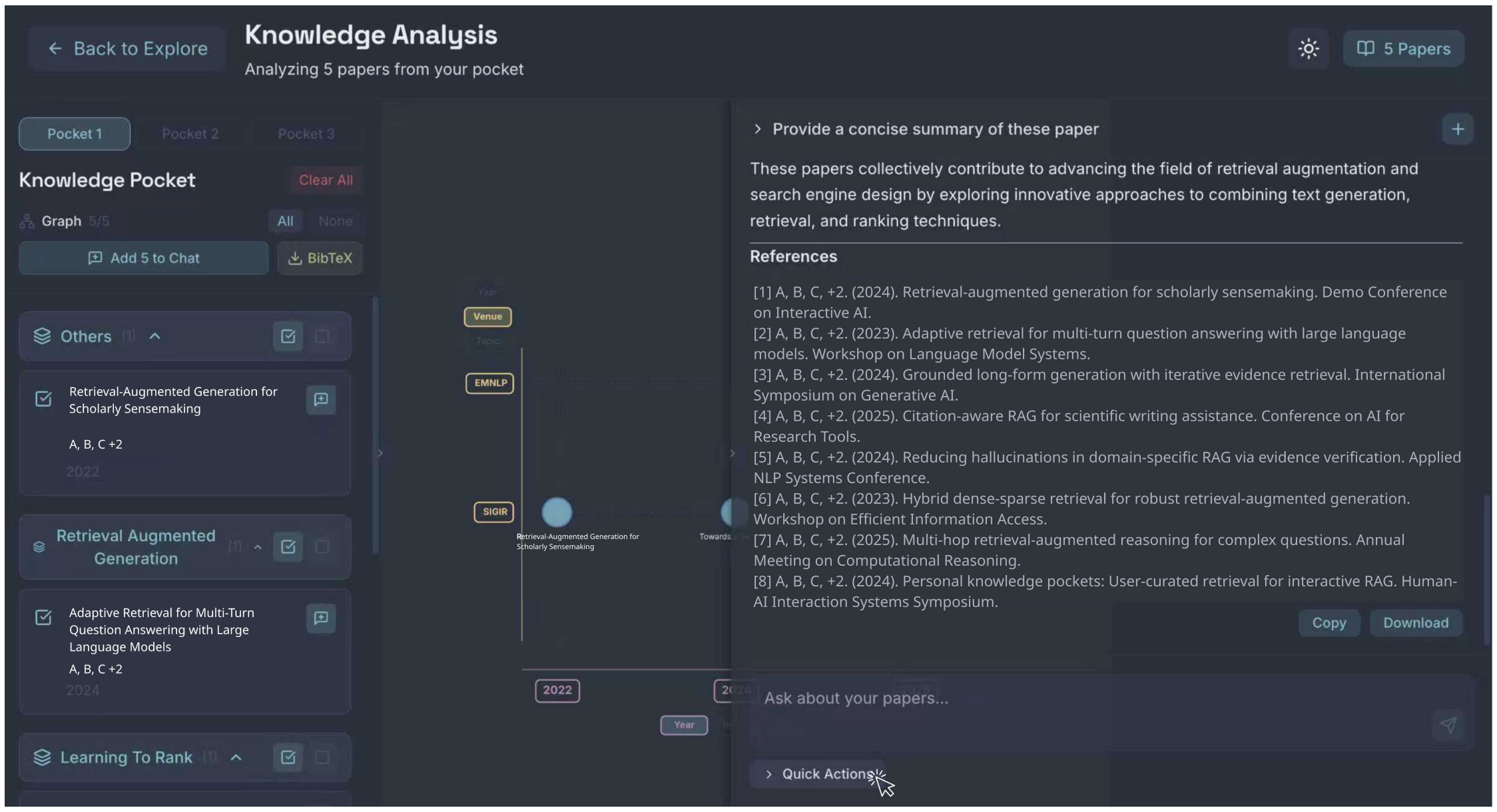}
    \caption{Prototype Synthesis Interface with Graph-Based Analysis and AI Assistant.}
    \label{fig:prototype5}
\end{figure*}

In this section, Figures~\ref{fig:prototype1}--\ref{fig:prototype5} present the core workflow of the system, including query input with AI-generated search suggestions, exploration of retrieved papers, paper recommendation, collection of relevant papers into knowledge pockets, and knowledge analysis supported by a LLM assistant. Together, these prototypes demonstrate the intended interaction flow and key functional components of the system.

%% file: appendix/data_process.tex
\section{\textsc{HiPaper} Behavior Log Processing Details}
\label{appendix:log}

This appendix provides the detailed preprocessing and computation procedures used to derive the two behavioral visualizations reported in the main paper: \textbf{Phase Progression Over Session Time} and  \textbf{Action-to-Action Transition Heatmap.}

\subsection{Data Preparation}
Interaction logs were stored as JSON files. Each session was assigned to one of two task conditions, \textit{Familiar} or \textit{Unfamiliar}, based on the study design. For all analyses, events within each session were sorted in ascending order by \texttt{timestamp\_ms}.

The two behavioral analyses used slightly different preprocessing steps because they captured different aspects of interaction. The Phase Progression Over Session Time analysis preserved the temporal structure of the full session, including events that did not map to one of the three analytical phases. In contrast, the Action-to-Action Transition analysis focused on meaningful action-to-action behavior and therefore excluded page-level navigation events before constructing transition matrices.

\begin{table*}[t]
\centering
\caption{Event mapping used across the two behavioral analyses. The phase-progression analysis mapped raw events to high-level sensemaking phases, while the transition analysis used human-readable action labels for visualization.}
\label{tab:event-mapping-all}
\begin{tabular}{p{0.24\textwidth} p{0.14\textwidth} p{0.22\textwidth} p{0.30\textwidth}}
\toprule
\textbf{Raw event type} & \textbf{Phase label} & \textbf{Heatmap label} & \textbf{Note} \\
\midrule
\texttt\texttt{search\_query} & Retrieval & Retrieval & User issues a search query to retrieve candidate papers or information in search page. \\
\texttt{header\_search} & Retrieval & Retrieval & User issues a search query to retrieve candidate papers or information in the header of explore page. \\
\texttt{action\_add\_to\_pocket} & Collection & Action Add To Pocket & User saves a paper or item to the pocket for later consideration. \\
\texttt{action\_dislike} & Collection & Dislike & User marks a paper or item as irrelevant or undesirable. \\
\texttt{action\_remove\_from\_pocket} & Collection & Action Remove From Pocket & User removes a previously saved paper or item from the pocket. \\
\texttt{action\_similar} & Collection & Similar & User requests or opens similar items related to the current paper. \\
\texttt{click} & Collection & Cluster & User clicks a clustered paper/item representation to inspect it. \\
\texttt{filter\_change} & Collection & Filter & User changes filtering criteria to refine the visible set of papers/items. \\
\texttt{ai\_insight\_add\_to\_chat} & Synthesis & AI Insight Add To Chat & User sends an AI-generated insight or suggested content into the chat workspace. \\
\texttt{ai\_insight\_node\_click} & Synthesis & AI Insight Node Click & User clicks an AI insight node to inspect or expand synthesized information. \\
\texttt{chat\_llm\_response} & Synthesis & Chat LLM Response & System-generated language-model response shown in the chat workspace. \\
\texttt{chat\_user\_message} & Synthesis & Chat User Message & User submits a message or prompt in the chat workspace. \\
\bottomrule
\end{tabular}
\end{table*}

\subsection{Phase Progression Over Session Time}
The Phase Progression analysis summarizes how participants distributed activity across the high-level phases of \textit{Retrieval}, \textit{Collection}, and \textit{Synthesis} over the session of a task.

\subsubsection{Session Normalization}
For each session, we identified the first and last event timestamps, denoted by \(t_0\) and \(t_1\). To make sessions of different durations comparable, we normalized each session onto a common progress scale from \(0\%\) to \(100\%\). For each event \(i\), its relative session position was computed as
\[
rel_i = \frac{timestamp\_ms_i - t_0}{t_1 - t_0} \times 100.
\]
Thus, \(0\%\) corresponds to the beginning of the session and \(100\%\) to the end, regardless of the session's absolute duration.

\subsubsection{Event-to-Phase Mapping}
Each event was mapped to one of four labels using a predefined phase mapping function: \textit{Retrieval}, \textit{Collection}, \textit{Synthesis}, or \textit{None}. The first three labels correspond to the high-level analytical phases used in \textsc{HiPaper}. Events that did not belong to any of these three categories were assigned to \textit{None}.  As a result, some portions of a session were intentionally left outside the three focal phases. The mapping details from the raw log to phase please refer to Table~\ref{tab:event-mapping-all}.

\subsubsection{Grid-Based Phase Assignment}
To derive a comparable session-level representation, we sampled each normalized session on a uniform grid of 401 points spanning \(0\%\) to \(100\%\) progress. For each grid point \(x\), we identified the event interval covering that point. Operationally, event \(i\) was treated as owning the interval from \(rel_i\) to \(rel_{i+1}\), and the phase label of event \(i\) was assigned to the full interval.

This procedure yielded, for each session, a binary indicator vector for each phase. At each grid point, the indicator takes value 1 if the session was in that phase at that point in normalized progress, and 0 otherwise.

\subsubsection{Aggregation Across Sessions}
Within each condition, we averaged the binary indicator vectors across all sessions. Therefore, for each phase and each grid point, the resulting value represents the proportion of sessions in that condition that were occupying that phase at that point in normalized progress.

This representation is dwell-weighted rather than event-count-weighted: if participants spent a long duration in a phase, that phase occupies a wider region of the plot even if relatively few events occurred during that time.

\subsubsection{Smoothing and Interpretation}
To reduce visual noise, the aggregated phase curves were smoothed using a Gaussian filter with \(\sigma = 3\) grid bins, corresponding to approximately \(0.75\%\) of normalized session progress. Smoothed values were clipped to the interval \([0,1]\).

An important consequence of the \textit{None} category is that the phase proportions at a given progress point do not necessarily sum to 1.0. Specifically, when sessions are in events that do not map to \textit{Retrieval}, \textit{Collection}, or \textit{Synthesis}, those sessions do not contribute to any of the three plotted phase proportions.

\subsection{Action-to-Action Transition analysis}
The transition analysis captures local action-to-action movement patterns within sessions and compares these patterns across the Familiar and Unfamiliar task conditions.

\subsubsection{Preprocessing}
Before constructing transitions, all events associate with page-level interaction were excluded (such as \texttt{event type=page change, label= to\_Collection}). These page-change events were not treated as meaningful interaction actions for this analysis. In addition, the event types \texttt{search\_query} and \texttt{header\_search} were merged into a single \texttt{retrieval} category to avoid splitting semantically equivalent search actions across multiple labels.

\subsubsection{Transition Extraction}
Within each \((participant\_id, session)\), events were sorted by \texttt{timestamp\_ms}. Each event was then paired with the immediately following event in the same session, producing a set of ordered transitions of the form \((current\_action, next\_action)\). Each transition inherited the condition label of its session.

\subsubsection{Transition Event Definitions}
For the transition analysis, rows with \texttt{category == "page"} were excluded before transition extraction. In addition, \texttt{header\_search} and \texttt{search\_query} were merged into a single \texttt{search} label. All other event types retained their original action labels. Table~\ref{tab:event-mapping-all} lists the action labels used in the transition matrix.

\subsubsection{Transition Count Matrices}
We grouped transitions by (condition, event\_type, next\_event) and counted their frequency. These counts were then pivoted into square transition matrices, one for each condition. Missing cells were filled with 0. To ensure direct comparability, both matrices used the same row and column ordering. Event labels were ordered first by broader stage grouping (\textit{Retrieval} \(\rightarrow\) \textit{Collection} \(\rightarrow\) \textit{Synthesis} \(\rightarrow\) \textit{Other}) and then alphabetically within each group.

\subsubsection{Row Normalization}
Because the two conditions contained different total numbers of transitions, direct comparison of raw counts would be influenced by exposure volume. To control for this, each transition matrix was row-normalized to obtain conditional probabilities:
\[
P(i,j) = \frac{Count(i,j)}{\sum_k Count(i,k)}.
\]
Equivalently, each row estimates the probability \(P(j \mid i)\), that is, the probability that action \(j\) occurs next given that the current action is \(i\).

If a row had zero total outgoing transitions in a condition, no conditional distribution could be computed for that action. In such cases, the row was treated as undefined (\texttt{NaN}) and masked in the heatmap visualization.

\subsubsection{Difference Matrix}
To compare conditions, we computed the difference between the two row-normalized matrices:
\[
\Delta(i,j) = P_{\mathrm{Fam}}(j \mid i) - P_{\mathrm{Unfam}}(j \mid i).
\]
Each cell therefore indicates how much more likely a given next action \(j\) was after current action \(i\) in familiar-topic sessions than in unfamiliar-topic sessions. Positive values indicate transitions favored by the Familiar condition, whereas negative values indicate transitions favored by the Unfamiliar condition.

Because each row of a row-normalized transition matrix sums to 1, each row of the difference matrix sums to 0, except for rows that are undefined due to missing outgoing transitions in one of the conditions.

\subsubsection{Interpretation}
This probability-difference formulation emphasizes relative transition tendencies rather than raw interaction volume. For example, a negative value for a self-transition such as \textit{Cluster} \(\rightarrow\) \textit{Cluster} indicates that, given a cluster action has just occurred, participants in the Unfamiliar condition were more likely than those in the Familiar condition to immediately perform another cluster action. This interpretation is independent of the total number of transitions observed in each condition.

\subsection{Summary of Analytical Differences}
The two behavioral visualizations differ in both their unit of analysis and their normalization strategy.

\begin{itemize}
    \item \textbf{Phase Progression Over Session Time} uses normalized session progress as the unit of analysis and preserves the temporal structure of the full session. It shows when participants were more likely to be in \textit{Retrieval}, \textit{Collection}, or \textit{Synthesis} over the course of a session.
    \item \textbf{Action-to-Action Transition Probability Heatmap} use event-to-event transitions as the unit of analysis. They show how the conditional probability of moving from one action to another differs between the Familiar and Unfamiliar conditions.
\end{itemize}

Together, these two analyses provide complementary perspectives on behavior: the Phase Progression Over Task Time captures macro-level temporal progression, while the transition-probability view captures micro-level interaction strategy.

%% file: appendix/dataset.tex
\section{Datasets}
\label{app:datasets}

\begin{table*}[ht]
\centering
\caption{Summary of Venues in the Database of \textsc{HiPaper} Systems.}
\label{tab:venue-summary}
\resizebox{\linewidth}{!}{
\begin{tabular}{llrllll}
\toprule
Abbreviation & Venue Name & \#Record & Domain & CORE & CCF & JCR \\
\midrule
AAAI & AAAI Conference on Artificial Intelligence & 11416 & Conference & A* & A & -- \\
AAMAS & International Conference on Autonomous Agents and Multiagent Systems & 2076 & Conference & A* & B & -- \\
ACL & Annual Meeting of the Association for Computational Linguistics & 8374 & Conference & A* & A & -- \\
CHI & ACM Conference on Human Factors in Computing Systems & 4612 & Conference & A* & A & -- \\
CSCW & ACM Conference on Computer-Supported Cooperative Work and Social Computing & 781 & Conference & A & A & -- \\
CVPR & IEEE/CVF Conference on Computer Vision and Pattern Recognition & 11674 & Conference & A* & A & -- \\
ECCV & European Conference on Computer Vision & 4034 & Conference & A* & B & -- \\
EMNLP & Conference on Empirical Methods in Natural Language Processing & 9230 & Conference & A* & B & -- \\
HCI &
Human–Computer Interaction & 104 & Journal & -- & -- & Q1 \\
ICCV & IEEE/CVF International Conference on Computer Vision & 4264 & Conference & A* & A & -- \\
ICDE & IEEE International Conference on Data Engineering & 1916 & Conference & A* & A & -- \\
ICLR & International Conference on Learning Representations & 9489 & Conference & A* & A & -- \\
ICML & International Conference on Machine Learning & 10186 & Conference & A* & A & -- \\
IEEE VR & IEEE Conference on Virtual Reality and 3D User Interfaces & 439 & Conference & A* & A & -- \\
IJCAI & International Joint Conference on Artificial Intelligence & 4757 & Conference & A* & B & -- \\
IJHCI & International Journal of Human-Computer Interaction & 2612 & Journal & -- & B & Q1 \\
IJHCS & International Journal of Human-Computer Studies & 597 & Journal & -- & A & Q1 \\
ISMAR & IEEE International Symposium on Mixed and Augmented Reality & 565 & Conference & A & B & -- \\
ISMAR-Adjunct & IEEE International Symposium on Mixed and Augmented Reality Adjunct & 973 & Conference & -- & -- & -- \\
JMLR & Journal of Machine Learning Research & 619 & Journal & -- & A & Q1 \\
KDD & ACM SIGKDD Conference on Knowledge Discovery and Data Mining & 2926 & Conference & A* & A & -- \\
MM & ACM International Conference on Multimedia & 2752 & Conference & A* & A & -- \\
NeurIPS & Conference on Neural Information Processing Systems & 13255 & Conference & A* & A & -- \\
SIGGRAPH & ACM SIGGRAPH Conference & 1112 & Conference & A* & A & -- \\
SIGGRAPH-ASIA & ACM SIGGRAPH Conference in Asia & 1053 & Conference & A* & A & -- \\
SIGIR & International ACM SIGIR Conference on Research and Development in Information Retrieval & 2179 & Conference & A* & A & -- \\
SIGMOD & ACM SIGMOD International Conference on Management of Data & 1157 & Conference & A* & A & -- \\
TKDE & IEEE Transactions on Knowledge and Data Engineering & 1206 & Journal & -- & A & Q1 \\
TMM & IEEE Transactions on Multimedia & 3071 & Journal & -- & A & Q1 \\
TNNLS & IEEE Transactions on Neural Networks and Learning Systems & 4708 & Journal & -- & B & Q1 \\
TOCHI & ACM Transactions on Computer-Human Interaction & 317 & Journal & -- & A & Q1 \\
TOG & ACM Transactions on Graphics & 1068 & Journal & -- & A & Q1 \\
TOIS & ACM Transactions on Information Systems & 552 & Journal & -- & A & Q1 \\
TVCG & IEEE Transactions on Visualization and Computer Graphics & 184 & Journal & -- & A & Q1 \\
UIST & ACM Symposium on User Interface Software and Technology & 1034 & Conference & A* & A & -- \\
VLDB & International Conference on Very Large Data Bases & 1924 & Conference & A* & A & -- \\
VRW & IEEE Conference on Virtual Reality Workshops & 1045 & Conference & -- & C & -- \\
WWW & The Web Conference & 3454 & Conference & A* & A & -- \\
\bottomrule
\end{tabular}}
\end{table*}

In our~\textsc{HiPaper}, we collected 131,715 papers published in 38 selected venues from 2021 to 2025. Table~\ref{tab:venue-summary} summarizes the 38 publication venues. For each venue, we report its abbreviation, full name, number of records, publication type, and commonly used venue quality indicators, including CORE, CCF, and JCR. The selected venues primarily cover major research areas relevant to our system, including HCI, AI, graphics, and data management.

Table~\ref{tab:data-schema} presents the schema of the paper records in our dataset. Each record is structured as a unified representation that integrates bibliographic metadata, content fields, and derived analytical attributes. The schema includes standard identifiers such as DOI, title, authors, venue, and year, as well as access information like PDF URLs and abstracts. In addition, we incorporate LLM-generated annotations, including high-level paper categories, concise TL;DR summaries of background and method, and hierarchical taxonomy labels. Citation-related statistics, such as yearly citation counts and percentile scores, are also included to support impact-aware analysis. This structured schema enables consistent indexing, retrieval, and synthesis over papers.

\begin{table*}[t]
\centering
\caption{Schema Summary of Paper Records (Examples are Anonymized).}
\label{tab:data-schema}
\resizebox{\linewidth}{!}{
\begin{tabular}{l l l l}
\hline
Key & Value Type & Example & Meaning \\
\hline
\texttt{doi} & \texttt{str} & \texttt{"10.1609/AAAI.V36I1.12345"} & DOI of the paper. \\
\texttt{title} & \texttt{str} & \texttt{"Anonymized Research Title on Learning Systems"} & Paper title. \\
\texttt{authors} & \texttt{list[str]} & \texttt{["Author A", "Author B", "Author C"]} & Ordered list of authors. \\
\texttt{venue} & \texttt{str} & \texttt{"AAAI"} & Venue abbreviation. \\
\texttt{year} & \texttt{int} / \texttt{str} & \texttt{2022} & Publication year.\\
\texttt{pdf\_url} & \texttt{str} & \texttt{"https://publisher.org/article/view/12345/12000"} & URL to the paper PDF. \\
\texttt{abstract} & \texttt{str} & \texttt{"This paper studies an anonymized problem..."} & Abstract text of the paper. \\
\texttt{paper\_category} & \texttt{str} & \texttt{"Methodology"} & High-level paper category. \\
\texttt{TL;DR (Background)} & \texttt{str} & \texttt{This paper address an anonymized problem in...} & Short background summary. \\
\texttt{TL;DR (Method)} & \texttt{str} & \texttt{This paper proposed an anonymized method...} & Short method summary. \\
\texttt{citation\_percentile} & \texttt{float} & \texttt{0.49} & Citation percentile score. \\
\texttt{cited\_by\_year} & \texttt{list[dict]} & \texttt{[\{"year": 2024, "cited\_by\_count": 1\}]} & Citation counts grouped by year. \\
\texttt{reference} & \texttt{list[str]} & \texttt{["https://openalex.org/W123...", ...]} & Referenced works or their identifiers. \\
\texttt{llm\_taxonomy} & \texttt{dict} & \texttt{\{"Level 1": "Machine Learning", "Level 2": ["Topic X"]\}} & LLM-generated taxonomy labels. \\
\hline
\end{tabular}}
\end{table*}

%% file: appendix/prompt.tex
\newcommand{\WC}[1]{\textcolor{lightblue}{\ttfamily\detokenize{<#1>}}}
\newcommand{\PSTR}[1]{{\ttfamily\detokenize{#1}}}
\definecolor{lightblue}{HTML}{4AA3DF}
\definecolor{titlegray}{HTML}{BFBFBF}
\definecolor{boxbg}{HTML}{F7F7F7}
\newtcolorbox{PromptStyleBox}[1]{
  enhanced,
  colback=boxbg,
  colframe=titlegray,
  boxrule=0.4pt,
  arc=2pt,
  boxsep=3pt,
  left=10pt,
  right=10pt,
  top=6pt,
  bottom=6pt,
  before skip=6pt,
  after skip=6pt,
  title=\textbf{#1},
  fonttitle=\bfseries,
  coltitle=black,
  boxed title style={
    colback=titlegray,
    colframe=titlegray,
    arc=0pt,
    left=8pt,
    right=8pt,
    top=4pt,
    bottom=4pt
  }
}
\newenvironment{PromptBody}
{%
  \ttfamily\small
  \parindent=0pt
  \parskip=0pt
  \raggedright
  \sloppy
  \hyphenpenalty=10000
  \exhyphenpenalty=10000
}
{\par}

\raggedbottom

\section{Prompts}
\label{app:prompts}
This section summarizes the prompts used across our \textsc{HiPaper}. They are organized by functional stage in our system, including Retrieval, Collection, and Synthesis.

\subsection{Retrieval Prompts}
In this section, we illustrate prompts used to convert raw user information needs into retrieval-friendly academic queries. These prompts improve semantic alignment between underspecified user requests and the language commonly used in paper titles and abstracts, thereby supporting more accurate vector-based paper retrieval.

\begin{figure*}[t]
\begin{PromptStyleBox}{Query Refinement Prompt}

\textbf{System Prompt:}

\begin{PromptBody}
You convert rough user search topics into compact academic retrieval queries for semantic vector search.

Produce a query that looks natural in research paper titles, abstracts, or author keywords, while staying faithful to the user's intent.

\textbf{Guidelines.}
\begin{itemize}[leftmargin=1.4em, label=-, itemsep=0pt]
  \item Preserve the original intent.
  \item Use natural academic phrasing to improve semantic alignment.
  \item Do not over-specify broad or underspecified queries with invented tasks, methods, domains, datasets, benchmarks, or constraints.
  \item Output only one query.
  \item Maximum 30 words.
\end{itemize}
\end{PromptBody}

\textbf{User Prompt:}

\begin{PromptBody}
\WC{raw_query}
\end{PromptBody}
\end{PromptStyleBox}
\end{figure*}

\subsection{Collection Prompts}
In this section, we enrich paper metadata via LLMs to enable a structured collection space. It supports hierarchical taxonomy assignment, taxonomy normalization, taxonomy merging, paper background and method summarization, and coarse paper-type identification, enabling consistent indexing and collection-level organization.

\begin{figure*}[t]
\begin{PromptStyleBox}{Level 1 Cluster Generation}

\textbf{System Prompt:}
\begin{PromptBody}
Classify the paper into exactly one existing subfield taxonomy. You must choose from the provided subfield list only. Do not create, rewrite, or paraphrase taxonomy names. Return JSON only in this format \{"subfield": "\WC{subfield}"\}.
\end{PromptBody}

\textbf{User Prompt:}
\begin{PromptBody}
\begin{itemize}[leftmargin=1.4em, label=-, itemsep=0pt]
  \item Paper title \WC{title}\\
  \item Paper abstract \WC{abstract}\\
  \item Existing subfield taxonomy \WC{subfield_choices_json_array}\\
  \item Rules
  \begin{itemize}[leftmargin=1.4em, label=-, itemsep=0pt]
    \item Output exactly one subfield taxonomy.
    \item The output must be exactly one item from Existing subfield taxonomy.
    \item The selected subfield must be the best fit for the paper.
  \end{itemize}
\end{itemize}
\end{PromptBody}
\end{PromptStyleBox}
\end{figure*}

\begin{figure*}[t]
\begin{PromptStyleBox}{Level 2 Cluster Generation}

\textbf{System Prompt:}

\begin{PromptBody}
Select exactly \WC{output_count} \WC{target_desc} taxonomy items for the paper. Strongly prefer reusing existing \WC{target_desc} taxonomy from the provided list. Only create a new taxonomy when none of the existing options is reasonably relevant. Return JSON only in this format \{"\WC{target_key}": ["taxonomy1", "taxonomy2", "..."]\}.
\end{PromptBody}

\textbf{User Prompt:}
\begin{PromptBody}
\begin{itemize}[leftmargin=1.4em, label=-, itemsep=0pt]
  \item Paper title \\
  \WC{title}
  \item Paper abstract \\
  \WC{abstract}
  \item Latest selected \WC{parent_key} \\
  \WC{parent_value}
  \item \WC{extra_context} Existing \WC{target_key} taxonomy \\
  \WC{target_choices_json_array}
  \item Rules
  \begin{itemize}[leftmargin=1.4em, label=-, itemsep=0pt]
    \item Output exactly \WC{output_count} unique \WC{target_key} items.
    \item All \WC{target_key} items must align with the selected \WC{parent_key}.
    \item Reuse existing taxonomy whenever possible, including partially related items.
    \item Create a new taxonomy item only if no existing item is reasonably relevant.
    \item New taxonomy must be broad, reusable, and without abbreviations.
    \item Each taxonomy item must be at most \WC{max_taxonomy_words} words.
    \item \WC{extra_generation_rules}
  \end{itemize}
  \item Previous failed output (fix issues and return valid JSON only): \\
  \WC{last_failed_response}
\end{itemize}
\end{PromptBody}
\end{PromptStyleBox}
\end{figure*}

\begin{figure*}[t]
\begin{PromptStyleBox}{Taxonomy Reflection}

\textbf{System Prompt:}

\begin{PromptBody}
You are a strict taxonomy normalizer for \WC{level_key}. Bias strongly toward reusing existing taxonomy terms. Choose covered=true by default when there is reasonable semantic overlap. Return JSON only: \{"covered": true, "taxonomy": "\WC{existing_taxonomy}"\} or \{"covered": false\}.
\end{PromptBody}

\textbf{User Prompt:}
\begin{PromptBody}
\begin{itemize}[leftmargin=1.4em, label=-, itemsep=0pt]
  \item Paper title \\
  \WC{title}
  \item Paper abstract \\
  \WC{abstract}
  \item Candidate new \WC{level_key} taxonomy \\
  \WC{candidate}
  \item Existing \WC{level_key} taxonomy list \\
  \WC{choices_json_array}
  \item Rules
  \begin{itemize}[leftmargin=1.4em, label=-, itemsep=0pt]
    \item Prefer covered=true whenever any existing taxonomy is semantically close, broader, narrower, or partially overlapping.
    \item Set covered=false only if the candidate is clearly unrelated to all existing taxonomy items.
    \item If covered=true, taxonomy must be exactly one item from the existing taxonomy list.
    \item Avoid creating new taxonomy unless absolutely necessary.
  \end{itemize}
  \item Previous failed output (fix issues and return valid JSON only): \\
  \WC{last_failed_response}
\end{itemize}
\end{PromptBody}
\end{PromptStyleBox}
\end{figure*}

\begin{figure*}[t]
\begin{PromptStyleBox}{Taxonomy Combination}

\textbf{System Prompt:}
\begin{PromptBody}
You merge semantically equivalent or highly overlapping taxonomy items into a unified set. Strongly prefer reusing existing taxonomy names. Do not create redundant or overly specific variants. Return JSON only in this format \WC{{"merged_taxonomy": ["taxonomy1", "taxonomy2", "..."]}}.
\end{PromptBody}

\textbf{User Prompt:}
\begin{PromptBody}
\begin{itemize}[leftmargin=1.4em, label=-, itemsep=0pt]
  \item Input taxonomy list \WC{taxonomy_list}\\
  \item Existing canonical taxonomy \WC{existing_taxonomy_list}\\
  \item Rules
  \begin{itemize}[leftmargin=1.4em, label=-, itemsep=0pt]
    \item Merge items with identical or highly similar meanings.
    \item Prefer a single representative term for each semantic group.
    \item Reuse existing taxonomy whenever possible.
    \item Avoid introducing new taxonomy unless necessary to resolve ambiguity.
    \item Ensure output taxonomy is minimal and non-redundant.
    \item Each taxonomy item must be concise and generalizable.
  \end{itemize}
\end{itemize}
\end{PromptBody}
\end{PromptStyleBox}
\end{figure*}

\begin{figure*}[t]
\begin{PromptStyleBox}{Paper Background and Method Summarization}

\textbf{System Prompt:}

\begin{PromptBody}
Write a TLDR in exactly two sentences. Sentence 1: research background/problem context. Sentence 2: method/approach. Each sentence must be <= 50 words. Return JSON only: \{"tldr": \{"background": "sentence1", "method": "sentence2"\}\}.
\end{PromptBody}

\textbf{User Prompt:}
\begin{PromptBody}
\begin{itemize}[leftmargin=1.4em, label=-, itemsep=0pt]
  \item Paper title \WC{title}\\
  \item Paper abstract \WC{abstract}\\
  \item Rules
  \begin{itemize}[leftmargin=1.4em, label=-, itemsep=0pt]
    \item Output exactly 2 sentences in order.
    \item Sentence 1 must describe background/problem context only.
    \item Sentence 2 must describe method/approach only.
    \item Each sentence must be at most \WC{max_words_per_sentence} words.
    \item Keep wording concise and specific.
  \end{itemize}
\end{itemize}
\end{PromptBody}
\end{PromptStyleBox}
\end{figure*}

\begin{figure*}[t]
\begin{PromptStyleBox}{Paper Type Infer}

\textbf{System Prompt:}

\begin{PromptBody}
You are a strict research paper category classifier. Given ONLY the title and abstract, determine the PRIMARY contribution category of the paper.

\textbf{Definitions.}
\begin{itemize}[leftmargin=1.4em, label=-, itemsep=0pt]
  \item \textbf{Methodology}. The paper proposes a novel model, algorithm, technical framework, or system. Experiments or datasets may be included, but the core contribution is a new technical method.
  \item \textbf{Benchmark}. The paper primarily introduces a new dataset, benchmark, evaluation protocol, or leaderboard. The main contribution is evaluation infrastructure rather than a new technical method.
  \item \textbf{Survey}. The paper systematically reviews, summarizes, or categorizes existing literature. It does NOT introduce a novel algorithm or technical framework.
  \item \textbf{Position Paper}. The paper presents a conceptual argument, research agenda, or vision without proposing a concrete technical method or systematic survey.
  \item \textbf{Workshop Proposal}. The paper proposes organizing a workshop or community event.
  \item \textbf{Others}. Anything that does not clearly fit the above categories.
\end{itemize}

Respond ONLY with a JSON object of the form: \WC{{"paper_category": "allowed_categories"}}.
\end{PromptBody}

\begin{PromptBody}
\textbf{User Prompt:}
\begin{itemize}[leftmargin=1.4em, label=-, itemsep=0pt]
  \item Title \WC{title} \\
  \item Abstract \WC{abstract} \\
\end{itemize}
\end{PromptBody}
\end{PromptStyleBox}
\end{figure*}

\subsection{Synthesis Prompts}
In this section, we prompt LLMs to synthesize information across multiple papers. They support paper relationship graph construction and validation, evidence extraction for question answering, answer drafting from selected evidence, and answer reflection for improving factual support, relevance, and clarity.

\begin{figure*}[t]
\begin{PromptStyleBox}{Paper Relationship Graph Generation}

\textbf{System Prompt:}

\begin{PromptBody}
You are an expert academic researcher and literature review assistant. Your task is to analyze a provided list of academic papers and generate a structured map of the relationships among them.

To do this, you will use the ``Entities and Relations'' framework and the typology of academic relationships developed by Shayan Doroudi (2023).

Typology:
\begin{itemize}[leftmargin=1.4em, label=-, itemsep=0pt]
  \item Intersection: Paper A and Paper B share common entities (concepts) or relations that are highly relevant to each other.
  \item Interpretation: Paper A provides a new lens or framework to interpret the findings of Paper B.
  \item Expansion: Paper B provides a missing link that expands the scope of Paper A (e.g., Paper A shows X causes Y; Paper B shows Y causes Z, expanding A to link X to Z).
  \item Abstraction: Paper B provides a generalized, theoretical framework that subsumes the specific, concrete findings of A.
  \item Reification: Paper B provides a concrete, specific example or historical instance of the abstract theory in A.
  \item Analogy: Papers A and B operate in different domains but share a deep structural similarity or underlying abstraction.
  \item Substitution: Paper A and Paper B have the exact same goal, but substitute one specific method for another.
  \item No\_meaningful\_relationship: The papers may share a trivial keyword, but they operate in entirely different contexts with no structural, causal, or thematic overlap worth noting.
\end{itemize}

Rules:
\begin{itemize}[leftmargin=1.4em, label=-, itemsep=0pt]
  \item Use source PaperId/targetPaperId exactly from provided ids.
  \item confidence must be numeric in [0, 1].
  \item evidence must be an array of short strings.
  \item Prefer no\_meaningful\_relationship instead of forcing weak links.
  \item Keep graph sparse and meaningful.
  \item Do not invent papers, fields, or IDs.
\end{itemize}
\WC{In-context learning cases}
\end{PromptBody}

~

\textbf{User Prompt:}

\begin{PromptBody}
Task query \\
\WC{query_or_default}

Papers \\
Paper 1: \WC{paper_1_id};  \WC{paper_1_title}; \WC{paper_1_venue}; \WC{paper_1_year}; \WC{paper_1_abstract}.

Paper 2: \WC{paper_2_id};  \WC{paper_2_title}; \WC{paper_2_venue}; \WC{paper_2_year}; \WC{paper_2_abstract}.

\WC{additional_papers_block}

Return JSON only.
\end{PromptBody}

\end{PromptStyleBox}
\end{figure*}

\begin{figure*}[t]
\begin{PromptStyleBox}{Paper Relationship Graph Validation}

\textbf{System Prompt:}

\begin{PromptBody}
You are a strict reviewer for an academic paper relationship graph, using the Doroudi (2023) typology. You receive the same paper list plus a draft graph. Your task is to review each edge and produce a corrected graph.

Output must be VALID JSON only, with this exact shape:
\begin{verbatim}
{
  "summary": "string",
  "edgesReviewed": [
    {
      "sourcePaperId": "string",
      "targetPaperId": "string",
      "relationType": "one allowed label",
      "confidence": 0.0,
      "reason": "short rationale referencing specific entities/relations",
      "evidence": ["short evidence snippet"],
      "validatorAction": "kept|relabeled|removed"
    }
  ]
}
\end{verbatim}

Rules:
\begin{itemize}[leftmargin=1.4em, label=-, itemsep=0pt]
  \item Keep only defensible edges.
  \item You may relabel or remove weak or incorrect edges.
  \item Use validatorAction to indicate review decision.
  \item If relationType is no\_meaningful\_relationship, validatorAction should usually be removed.
  \item confidence must be numeric in [0, 1].
  \item Do not add IDs not present in the provided papers.
  \item JSON only. No markdown.
\end{itemize}
\end{PromptBody}

\textbf{User Prompt:}

\begin{PromptBody}
Task query: \WC{query_or_default}

Papers \\
Paper 1: \WC{paper_1_id};  \WC{paper_1_title}; \WC{paper_1_venue}; \WC{paper_1_year}; \WC{paper_1_abstract}.

Paper 2: \WC{paper_2_id};  \WC{paper_2_title}; \WC{paper_2_venue}; \WC{paper_2_year}; \WC{paper_2_abstract}.

\WC{additional_papers_block}

Draft graph JSON to review \\
\WC{draft_graph_json}

Return JSON only.
\end{PromptBody}

\end{PromptStyleBox}
\end{figure*}

\begin{figure*}[t]
\begin{PromptStyleBox}{Evidence Extract}

\textbf{System Prompt:}

\begin{PromptBody}
You are a research extraction agent. Given a set of papers and a question, extract the most relevant evidence sentences from each paper's abstract. Output ONLY a plain list, one line per evidence. Each line must start with the paper tag followed by a colon and the evidence sentence. Example format:

[1]: Evidence sentence from paper 1.

[2]: Evidence sentence from paper 2.

[1]: Another evidence from paper 1.

Do not add any headers, bullets, markdown formatting, or extra text.
\end{PromptBody}

\textbf{User Prompt:}

\begin{PromptBody}
Question \\
\WC{question}

Papers \\
\WC{paper_contexts_joined}
\end{PromptBody}

\end{PromptStyleBox}
\end{figure*}

\begin{figure*}[t]
\begin{PromptStyleBox}{Answer Drafting}

\textbf{System Prompt:}

\begin{PromptBody}
You are a research assistant. Based on the extracted evidence below, write a well-structured answer to the user's question. Use inline citation tags like [1], [2] to reference specific papers. Only cite papers that are directly relevant to your answer --- do NOT cite every paper. Do NOT include a References section at the end --- it will be appended automatically. Be concise and accurate.
\end{PromptBody}

\textbf{User Prompt:}

\begin{PromptBody}
Question \\
\WC{question}

Extracted evidence \\
\WC{extracted_evidence}
\end{PromptBody}

\end{PromptStyleBox}
\end{figure*}

\begin{figure*}[t]
\begin{PromptStyleBox}{Answer Reflection}

\textbf{System Prompt:}

\begin{PromptBody}
You are an answer reflection agent. Review the drafted answer against the user question and the extracted evidence. Check whether the answer is accurate, complete, concise, and fully supported by the cited evidence. Remove unsupported claims, fix misinterpretations, and improve relevance and structure. Preserve inline citation tags like [1], [2]. Return JSON only in this format \{"answer": "revised\_answer"\}.
\end{PromptBody}

\textbf{User Prompt:}

\begin{PromptBody}
Question \\
\WC{question}

Extracted evidence \\
\WC{extracted_evidence}

Draft answer \\
\WC{draft_answer}

Rules
\begin{itemize}[leftmargin=1.4em, label=-, itemsep=0pt]
  \item Ensure every substantive claim is supported by the extracted evidence.
  \item Remove claims that are unsupported, weakly supported, or irrelevant to the question.
  \item Keep only citations that directly support the associated statement.
  \item Improve clarity, logical flow, and conciseness.
  \item Do not add a references section.
  \item Preserve the original meaning unless correction is required by the evidence.
\end{itemize}

Previous failed output (fix issues and return valid JSON only) \\
\WC{last_failed_response}
\end{PromptBody}

\end{PromptStyleBox}
\end{figure*}

%% file: bib.bib
@String{Computing = "Computing" }

@String{Computer = "{IEEE} Computer" }

@String{Academic = "Academic Press" }

@String{Springer = "Springer-Verlag" }

@article{2024Urquhart_Sense_making,
author = {Urquhart, Christine and Cheuk, Bonnie and Lam, Louisa and Snowden, Dave},
title = {Sense‐making, sensemaking and sense making—A systematic review and meta‐synthesis of literature in information science and education: An Annual Review of Information Science and Technology (ARIST) paper},
year = {2024},
issue_date = {January 2025},
publisher = {John Wiley \& Sons, Inc.},
address = {USA},
volume = {76},
number = {1},
issn = {2330-1635},
url = {https://doi.org/10.1002/asi.24866},
doi = {10.1002/asi.24866},
journal = {J. Assoc. Inf. Sci. Technol.},
month = dec,
pages = {3–97},
numpages = {95}
}

@inproceedings{ma2025garden,
  title={Garden of Papers: Finding, Reading, and Organizing Research Papers in a Visual, Integrated, and Flexible Workspace},
  author={Ma, Donghyeok and Jang, Hanbee and Lee, Joon Hyub and Bae, Seok-Hyung},
  booktitle={Proceedings of the 38th Annual ACM Symposium on User Interface Software and Technology},
  pages={1--15},
  year={2025}
}

@article{NIGHTINGALE2009381,
title = {A guide to systematic literature reviews},
journal = {Surgery (Oxford)},
volume = {27},
number = {9},
pages = {381-384},
year = {2009},
note = {Determining surgical efficacy},
issn = {0263-9319},
doi = {https://doi.org/10.1016/j.mpsur.2009.07.005},
url = {https://www.sciencedirect.com/science/article/pii/S0263931909001707},
author = {Alison Nightingale},
}

@article{Whitfield03072023,
author = {Sharon Whitfield and Melissa A. Hofmann},
title = {Elicit: AI literature review research assistant},
journal = {Public Services Quarterly},
volume = {19},
number = {3},
pages = {201--207},
year = {2023},
publisher = {Routledge},
doi = {10.1080/15228959.2023.2224125},
URL = {

        https://doi.org/10.1080/15228959.2023.2224125

},
eprint = {

        https://doi.org/10.1080/15228959.2023.2224125

}

}

@InProceedings{10.1007/978-3-031-56069-9_28,
author="Jain, Siddhant
and Kumar, Asheesh
and Roy, Trinita
and Shinde, Kartik
and Vignesh, Goutham
and Tondulkar, Rohan",
editor="Goharian, Nazli
and Tonellotto, Nicola
and He, Yulan
and Lipani, Aldo
and McDonald, Graham
and Macdonald, Craig
and Ounis, Iadh",
title="SciSpace Literature Review: Harnessing AI for Effortless Scientific Discovery",
booktitle="Advances in Information Retrieval",
year="2024",
publisher="Springer Nature Switzerland",
address="Cham",
pages="256--260",
isbn="978-3-031-56069-9"
}

@inproceedings{horvitz1999principles,
  title={Principles of mixed-initiative user interfaces},
  author={Horvitz, Eric},
  booktitle={Proceedings of the SIGCHI conference on Human Factors in Computing Systems},
  pages={159--166},
  year={1999}
}

@inproceedings{amershi2019guidelines,
  title={Guidelines for human-AI interaction},
  author={Amershi, Saleema and Weld, Dan and Vorvoreanu, Mihaela and Fourney, Adam and Nushi, Besmira and Collisson, Penny and Suh, Jina and Iqbal, Shamsi and Bennett, Paul N and Inkpen, Kori and others},
  booktitle={Proceedings of the 2019 chi conference on human factors in computing systems},
  pages={1--13},
  year={2019}
}

@article{bittle2025generative,
  title={Generative AI and academic integrity in higher education: A systematic review and research agenda},
  author={Bittle, Kyle and El-Gayar, Omar},
  journal={Information},
  volume={16},
  number={4},
  pages={296},
  year={2025},
  publisher={MDPI}
}

@inproceedings{10.1145/3707640.3731913,
author = {Ye, Runlong and Lee, Patrick and Varona, Matthew and Huang, Oliver and Nobre, Carolina},
title = {ScholarMate: A Mixed-Initiative Tool for Qualitative Knowledge Work and Information Sensemaking},
year = {2025},
isbn = {9798400713972},
publisher = {Association for Computing Machinery},
address = {New York, NY, USA},
url = {https://doi.org/10.1145/3707640.3731913},
doi = {10.1145/3707640.3731913},
booktitle = {Adjunct Proceedings of the 4th Annual Symposium on Human-Computer Interaction for Work},
articleno = {7},
numpages = {7},
location = {
},
series = {CHIWORK '25 Adjunct}
}

@article{liu2022frontier,
  title={Frontier and hot topics in electrochemiluminescence sensing technology based on CiteSpace bibliometric analysis},
  author={Liu, Xiancheng and Zhao, Shuang and Tan, Lili and Tan, Yu and Wang, Yi and Ye, Zhiyi and Hou, Changjun and Xu, Yan and Liu, Shan and Wang, Guixue},
  journal={Biosensors and Bioelectronics},
  volume={201},
  pages={113932},
  year={2022},
  publisher={Elsevier}
}

@article{shah2020prosumption,
  title={Prosumption: bibliometric analysis using HistCite and VOSviewer},
  author={Shah, Syed Hamad Hassan and Lei, Shen and Ali, Muhammad and Doronin, Dmitrii and Hussain, Syed Talib},
  journal={Kybernetes},
  volume={49},
  number={3},
  pages={1020--1045},
  year={2020},
  publisher={Emerald Publishing Limited}
}

@inproceedings{suh2023sensecape,
  title={Sensecape: Enabling multilevel exploration and sensemaking with large language models},
  author={Suh, Sangho and Min, Bryan and Palani, Srishti and Xia, Haijun},
  booktitle={Proceedings of the 36th annual ACM symposium on user interface software and technology},
  pages={1--18},
  year={2023}
}

@inproceedings{zhang2025paperbridge,
  title={PaperBridge: Crafting Research Narratives through Human-AI Co-Exploration},
  author={Zhang, Runhua and Ouyang, Yang and Shen, Leixian and Tang, Yuying and Ma, Xiaojuan and Qu, Huamin and Xu, Xian},
  booktitle={Proceedings of the 38th Annual ACM Symposium on User Interface Software and Technology},
  pages={1--21},
  year={2025}
}

@article{allabadi2025improving,
  title={Improving and Evaluating Open Deep Research Agents},
  author={Allabadi, Doaa and Bradbury, Kyle and Malof, Jordan M},
  journal={arXiv preprint arXiv:2508.10152},
  year={2025}
}

@article{gnewuch2022opposing,
  title={Opposing Effects of Response Time in Human--Chatbot Interaction: U. Gnewuch et al.},
  author={Gnewuch, Ulrich and Morana, Stefan and Adam, Marc TP and Maedche, Alexander},
  journal={Business \& Information Systems Engineering},
  volume={64},
  number={6},
  pages={773--791},
  year={2022},
  publisher={Springer}
}

@article{bergstrom2022sense,
  title={Sense of agency and user experience: Is there a link?},
  author={Bergstr{\"o}m, Joanna and Knibbe, Jarrod and Pohl, Henning and Hornb{\ae}k, Kasper},
  journal={ACM Transactions on Computer-Human Interaction (TOCHI)},
  volume={29},
  number={4},
  pages={1--22},
  year={2022},
  publisher={ACM New York, NY}
}

@article{bickmore2005establishing,
  title={Establishing and maintaining long-term human-computer relationships},
  author={Bickmore, Timothy W and Picard, Rosalind W},
  journal={ACM Transactions on Computer-Human Interaction (TOCHI)},
  volume={12},
  number={2},
  pages={293--327},
  year={2005},
  publisher={ACM New York, NY, USA}
}

@article{jansen2025human,
  title={Human-in-the-Loop Optimization for Inclusive Design: Balancing Automation and Designer Expertise},
  author={Jansen, Pascal},
  journal={arXiv preprint arXiv:2505.08375},
  year={2025}
}

@article{Natarajan2025,
  author = {Sriraam Natarajan and Saurabh Mathur 0002 and Sahil Sidheekh and Wolfgang Stammer and Kristian Kersting},
  title = {Human-in-the-loop or AI-in-the-loop? Automate or Collaborate?},
  journal = {AAAI},
  year = {2025},
  url = {https://ojs.aaai.org/index.php/AAAI/article/view/35083/37238},
}

@article{cowan2010magical,
  title={The magical mystery four: How is working memory capacity limited, and why?},
  author={Cowan, Nelson},
  journal={Current directions in psychological science},
  volume={19},
  number={1},
  pages={51--57},
  year={2010},
  publisher={Sage Publications Sage CA: Los Angeles, CA}
}

@inproceedings{mackenzie1992extending,
  title={Extending Fitts' law to two-dimensional tasks},
  author={MacKenzie, I Scott and Buxton, William},
  booktitle={Proceedings of the SIGCHI conference on Human factors in computing systems},
  pages={219--226},
  year={1992}
}

@article{doroudi2023related,
  title={What is a related work? A typology of relationships in research literature},
  author={Doroudi, Shayan},
  journal={Synthese},
  volume={201},
  number={1},
  pages={24},
  year={2023},
  publisher={Springer}
}

@article{brooke1996sus,
  title={SUS-A quick and dirty usability scale},
  author={Brooke, John and others},
  journal={Usability evaluation in industry},
  volume={189},
  number={194},
  pages={4--7},
  year={1996},
  publisher={London, England}
}

@inproceedings{hart2006nasa,
  title={NASA-task load index (NASA-TLX); 20 years later},
  author={Hart, Sandra G},
  booktitle={Proceedings of the human factors and ergonomics society annual meeting},
  volume={50},
  number={9},
  pages={904--908},
  year={2006},
  organization={Sage publications Sage CA: Los Angeles, CA}
}

@article{marangunic2015technology,
  title={Technology acceptance model: a literature review from 1986 to 2013},
  author={Maranguni{\'c}, Nikola and Grani{\'c}, Andrina},
  journal={Universal access in the information society},
  volume={14},
  number={1},
  pages={81--95},
  year={2015},
  publisher={Springer}
}

@article{marchionini2006exploratory,
  title={Exploratory search: from finding to understanding},
  author={Marchionini, Gary},
  journal={Communications of the ACM},
  volume={49},
  number={4},
  pages={41--46},
  year={2006},
  publisher={ACM New York, NY, USA}
}

@inproceedings{pirolli2005sensemaking,
  title={The sensemaking process and leverage points for analyst technology as identified through cognitive task analysis},
  author={Pirolli, Peter and Card, Stuart},
  booktitle={Proceedings of international conference on intelligence analysis},
  volume={5},
  number={1},
  pages={2--4},
  year={2005},
  organization={McLean, VA, USA}
}

@inproceedings{russell1993cost,
  title={The cost structure of sensemaking},
  author={Russell, Daniel M and Stefik, Mark J and Pirolli, Peter and Card, Stuart K},
  booktitle={Proceedings of the INTERACT'93 and CHI'93 conference on Human factors in computing systems},
  pages={269--276},
  year={1993}
}

@article{shneiderman2020human,
  title={Human-centered artificial intelligence: Reliable, safe \& trustworthy},
  author={Shneiderman, Ben},
  journal={International Journal of Human--Computer Interaction},
  volume={36},
  number={6},
  pages={495--504},
  year={2020},
  publisher={Taylor \& Francis}
}

@article{uren2006sensemaking,
  title={Sensemaking tools for understanding research literatures: Design, implementation and user evaluation},
  author={Uren, Victoria and Shum, Simon Buckingham and Bachler, Michelle and Li, Gangmin},
  journal={International journal of human-computer studies},
  volume={64},
  number={5},
  pages={420--445},
  year={2006},
  publisher={Elsevier}
}

@article{pontis2016understanding,
  title={Understanding “influence”: An empirical test of the D ata-F rame T heory of S ensemaking},
  author={Pontis, Sheila and Blandford, Ann},
  journal={Journal of the Association for Information Science and Technology},
  volume={67},
  number={4},
  pages={841--858},
  year={2016},
  publisher={Wiley Online Library}
}

@article{knopf2006doing,
  title={Doing a literature review},
  author={Knopf, Jeffrey W},
  journal={PS: Political Science \& Politics},
  volume={39},
  number={1},
  pages={127--132},
  year={2006},
  publisher={Cambridge University Press}
}

@article{farooq2017framework,
  title={A framework for identifying research gap in social sciences: Evidence from the past},
  author={Farooq, Rayees},
  journal={IUP Journal of Management Research},
  volume={16},
  number={4},
  pages={66--75},
  year={2017},
  publisher={IUP Publications}
}

@article{nicholson2018introduction,
  title={What do introduction sections tell us about the intent of scholarly work: A contribution on contributions},
  author={Nicholson, John D and LaPlaca, Peter and Al-Abdin, Ahmed and Breese, Richard and Khan, Zaheer},
  journal={Industrial Marketing Management},
  volume={73},
  pages={206--219},
  year={2018},
  publisher={Elsevier}
}

@article{van2024literature,
  title={Literature review papers: the search and selection process},
  author={van Wee, Bert and Banister, David},
  journal={Journal of Decision Systems},
  volume={33},
  number={4},
  pages={559--565},
  year={2024},
  publisher={Taylor \& Francis}
}

@article{yamada2025ai,
  title={The ai scientist-v2: Workshop-level automated scientific discovery via agentic tree search},
  author={Yamada, Yutaro and Lange, Robert Tjarko and Lu, Cong and Hu, Shengran and Lu, Chris and Foerster, Jakob and Clune, Jeff and Ha, David},
  journal={arXiv preprint arXiv:2504.08066},
  year={2025}
}

@article{eger2025transforming,
  title={Transforming science with large language models: A survey on ai-assisted scientific discovery, experimentation, content generation, and evaluation},
  author={Eger, Steffen and Cao, Yong and D'Souza, Jennifer and Geiger, Andreas and Greisinger, Christian and Gross, Stephanie and Hou, Yufang and Krenn, Brigitte and Lauscher, Anne and Li, Yizhi and others},
  journal={arXiv preprint arXiv:2502.05151},
  year={2025}
}

@article{ren2025towards,
  title={Towards scientific intelligence: A survey of llm-based scientific agents},
  author={Ren, Shuo and Xie, Can and Jian, Pu and Ren, Zhenjiang and Leng, Chunlin and Zhang, Jiajun},
  journal={arXiv preprint arXiv:2503.24047},
  year={2025}
}

@inproceedings{reddy2025towards,
  title={Towards scientific discovery with generative ai: Progress, opportunities, and challenges},
  author={Reddy, Chandan K and Shojaee, Parshin},
  booktitle={Proceedings of the AAAI conference on artificial intelligence},
  volume={39},
  number={27},
  pages={28601--28609},
  year={2025}
}

@article{cai2025impact,
  title={Impact of GPT on the Academic Ecosystem},
  author={Cai, Yudong and Deng, Qiyue and Lv, Ting and Zhang, Wan and Zhou, Yi},
  journal={Science \& Education},
  volume={34},
  number={2},
  pages={913--931},
  year={2025},
  publisher={Springer}
}

@inproceedings{he2025pasa,
  title={Pasa: An llm agent for comprehensive academic paper search},
  author={He, Yichen and Huang, Guanhua and Feng, Peiyuan and Lin, Yuan and Zhang, Yuchen and Li, Hang and others},
  booktitle={Proceedings of the 63rd Annual Meeting of the Association for Computational Linguistics (Volume 1: Long Papers)},
  pages={11663--11679},
  year={2025}
}

@article{jordan2025ranking,
  title={Ranking ‘by relevance’in academic literature searches: Prevalence, definitions, and implications},
  author={Jordan, Katy and Tsai, Sally Po},
  journal={Postdigital Science and Education},
  volume={7},
  number={4},
  pages={1092--1106},
  year={2025},
  publisher={Springer}
}

@article{fahrudin2025improved,
  title={An improved reference paper collection system using web scraping with three enhancements},
  author={Fahrudin, Tresna Maulana and Funabiki, Nobuo and Brata, Komang Candra and Naing, Inzali and Aung, Soe Thandar and Muhaimin, Amri and Prasetya, Dwi Arman},
  journal={Future Internet},
  volume={17},
  number={5},
  pages={195},
  year={2025},
  publisher={MDPI}
}

@inproceedings{ley2002dblp,
  title={The DBLP computer science bibliography: Evolution, research issues, perspectives},
  author={Ley, Michael},
  booktitle={International symposium on string processing and information retrieval},
  pages={1--10},
  year={2002},
  organization={Springer}
}

@article{priem2022openalex,
  title={OpenAlex: A fully-open index of scholarly works, authors, venues, institutions, and concepts},
  author={Priem, Jason and Piwowar, Heather and Orr, Richard},
  journal={arXiv preprint arXiv:2205.01833},
  year={2022}
}

@misc{jina_embeddings_v5,
  title={jina-embeddings-v5: General-purpose text embeddings},
  author={Jina AI},
  year={2024},
  url={https://jina.ai/embeddings/}
}

@misc{qdrant,
  title={Qdrant: Vector Database for the Next Generation of AI Applications},
  author={Qdrant Team},
  year={2023},
  url={https://qdrant.tech}
}

@misc{acm_ccs,
  title={ACM Computing Classification System},
  author={Association for Computing Machinery},
  year={2012},
  url={https://dl.acm.org/ccs}
}

@article{yang2025qwen3,
  title={Qwen3 technical report},
  author={Yang, An and Li, Anfeng and Yang, Baosong and Zhang, Beichen and Hui, Binyuan and Zheng, Bo and Yu, Bowen and Gao, Chang and Huang, Chengen and Lv, Chenxu and others},
  journal={arXiv preprint arXiv:2505.09388},
  year={2025}
}

@misc{gpt5,
  title={GPT-5 System Card},
  author={OpenAI},
  year={2025},
  url={https://openai.com}
}

@misc{arxiv,
  title        = {arXiv.org e-Print archive},
  author       = {{Cornell University}},
  year         = {2000},
  howpublished = {\url{https://arxiv.org}}
}

@misc{openreview,
  title        = {OpenReview},
  author       = {{OpenReview}},
  year         = {2018},
  howpublished = {\url{https://openreview.net}}
}

@article{alfaruque2023integrating,
  title={Integrating literature with technology and use of digital tools: Impact on learning outcomes},
  author={Alfaruque, Sabina Yasmin and Sultana, Sufia and Rastogi, Richa and Jabeen, Zakira},
  journal={World},
  volume={13},
  number={1},
  year={2023}
}

@misc{googlescholar,
  title        = {Google Scholar},
  author       = {{Google}},
  year         = {2004},
  howpublished = {\url{https://scholar.google.com/}}
}

@misc{semanticscholar,
  title        = {Semantic Scholar},
  author       = {{Semantic Scholar}},
  year         = {2015},
  howpublished = {\url{https://www.semanticscholar.org/}}
}

@misc{researchrabbit,
  title        = {Research Rabbit},
  author       = {{Research Rabbit}},
  year         = {2021},
  howpublished = {\url{https://www.researchrabbit.ai/}}
}

@article{kacperski2023examining,
  title={Examining bias perpetuation in academic search engines: An algorithm audit of Google and Semantic Scholar},
  author={Kacperski, Celina and Bielig, Mona and Makhortykh, Mykola and Sydorova, Maryna and Ulloa, Roberto},
  journal={arXiv preprint arXiv:2311.09969},
  year={2023}
}

@inproceedings{jiang2025hibench,
  title={Hibench: Benchmarking llms capability on hierarchical structure reasoning},
  author={Jiang, Zhuohang and Wu, Pangjing and Liang, Ziran and Chen, Peter Q and Yuan, Xu and Jia, Ye and Tu, Jiancheng and Li, Chen and Ng, Peter HF and Li, Qing},
  booktitle={Proceedings of the 31st ACM SIGKDD Conference on Knowledge Discovery and Data Mining V. 2},
  pages={5505--5515},
  year={2025}
}

@article{mei2001use,
  title={Use of filter paper for the collection and analysis of human whole blood specimens},
  author={Mei, Joanne V and Alexander, J Richard and Adam, Barbara W and Hannon, W Harry},
  journal={The Journal of nutrition},
  volume={131},
  number={5},
  pages={1631S--1636S},
  year={2001},
  publisher={Oxford University Press}
}

@article{wilcox2012research,
  title={Research data collection methods: from paper to tablet computers},
  author={Wilcox, Adam B and Gallagher, Kathleen D and Boden-Albala, Bernadette and Bakken, Suzanne R},
  journal={Medical care},
  volume={50},
  pages={S68--S73},
  year={2012},
  publisher={LWW}
}

@article{zhang2012comparison,
  title={Comparison of select reference management tools},
  author={Zhang, Yingting},
  journal={Medical reference services quarterly},
  volume={31},
  number={1},
  pages={45--60},
  year={2012},
  publisher={Taylor \& Francis}
}

@misc{endnote,
  title        = {EndNote},
  author       = {{EndNote}},
  year         = {1989},
  howpublished = {\url{https://endnote.com/}}
}

@misc{zotero,
  title        = {Zotero},
  author       = {{Zotero}},
  year         = {2006},
  howpublished = {\url{https://www.zotero.org/}}
}

@article{schick2016synthesis,
  title={What synthesis methodology should I use? A review and analysis of approaches to research synthesis},
  author={Schick-Makaroff, Kara and MacDonald, Marjorie and Plummer, Marilyn and Burgess, Judy and Neander, Wendy},
  journal={AIMS public health},
  volume={3},
  number={1},
  pages={172},
  year={2016}
}

@article{suri2009advancements,
  title={Advancements in research synthesis methods: From a methodologically inclusive perspective},
  author={Suri, Harsh and Clarke, David},
  journal={Review of Educational Research},
  volume={79},
  number={1},
  pages={395--430},
  year={2009},
  publisher={Sage Publications Sage CA: Los Angeles, CA}
}

@article{zhao2023survey,
  title={A survey of large language models},
  author={Zhao, Wayne Xin and Zhou, Kun and Li, Junyi and Tang, Tianyi and Wang, Xiaolei and Hou, Yupeng and Min, Yingqian and Zhang, Beichen and Zhang, Junjie and Dong, Zican and others},
  journal={arXiv preprint arXiv:2303.18223},
  volume={1},
  number={2},
  pages={1--124},
  year={2023}
}

@article{ellis1998paradigms,
  title={Paradigms and research traditions in information retrieval research},
  author={Ellis, David},
  journal={Information Services and Use},
  volume={18},
  number={4},
  pages={225--241},
  year={1998},
  publisher={SAGE Publications Sage UK: London, England}
}

@article{eimaraghy2004assessment,
  title={Assessment of manufacturing operational complexity},
  author={EIMaraghy, WH and Urbanic, RJ},
  journal={CIRP annals},
  volume={53},
  number={1},
  pages={401--406},
  year={2004},
  publisher={Elsevier}
}

@article{rivera2007measuring,
  title={Measuring the impact of Lean tools on the cost--time investment of a product using cost--time profiles},
  author={Rivera, Leonardo and Chen, F Frank},
  journal={Robotics and Computer-Integrated Manufacturing},
  volume={23},
  number={6},
  pages={684--689},
  year={2007},
  publisher={Elsevier}
}

@article{touvron2023llama,
  title={Llama: Open and efficient foundation language models},
  author={Touvron, Hugo and Lavril, Thibaut and Izacard, Gautier and Martinet, Xavier and Lachaux, Marie-Anne and Lacroix, Timoth{\'e}e and Rozi{\`e}re, Baptiste and Goyal, Naman and Hambro, Eric and Azhar, Faisal and others},
  journal={arXiv preprint arXiv:2302.13971},
  year={2023}
}
